\documentclass[a4paper]{aa}
\usepackage{graphicx}
\def\spose#1{\hbox to 0pt{#1\hss}}
\def\lta{\mathrel{\spose{\lower 3pt\hbox{$\mathchar"218$}}
     \raise 2.0pt\hbox{$\mathchar"13C$}}}
\def\gta{\mathrel{\spose{\lower 3pt\hbox{$\mathchar"218$}}
     \raise 2.0pt\hbox{$\mathchar"13E$}}}
\def\half{\frac{1}{2}}
\def\third{\frac{1}{3}}
\def\etal{{\it et al.\ }}
\begin{document}
   \thesaurus{}
  \title{Young Star Clusters in The Antennae: A Clue to their Nature from 
  	Evolutionary Synthesis}      


%
%
  \author{ Uta Fritze $-$ v. Alvensleben}
%
%
  \institute{ Universit\"atssternwarte G\"ottingen\\
  		 Geismarlandstr. 11, D - 37083  G\"ottingen, Germany\\
              email: ufritze@uni-sw.gwdg.de
             }
  \thesaurus{11.19.4, 11.09.1, 11.09.2, 11.09.3}
  \titlerunning{Young Star Clusters in the Antennae}

  \date{} 

\maketitle

\begin{abstract}

We analyse the population of bright star clusters in the interacting galaxy 
pair NGC 4038/39 detected with HST WFPC1 by 
Whitmore \& Schweizer (1995). 
Making use of our spectrophotometric evolutionary synthesis models for 
various initial metallicities we derive the ages of these star clusters 
and calculate their future luminosity evolution. This allows us to compare 
their luminosity function ({\bf LF}), evolved over a Hubble time, to LFs 
observed for the Milky Way's and other galaxies' star cluster systems. 
Since effective radii are difficult to determine due to 
crowding of the clusters, the shape of the LF after a Hubble time may 
help decide whether the young clusters are young globular clusters ({\bf GC}) or 
rather 
open clusters/OB associations. 
We find an intriguing difference in the shapes of the LFs if we subdivide the 
cluster population into subsamples with small and large effective radii. 
While the LF for the extended clusters looks exponential, that for clusters 
with small effective radii clearly shows a turn-over brighter than the 
completeness limit. For other possible subdivisions as to luminosity or colour 
no comparable differences are found. Evolving, in a first step, the LF from 
a common mean age of 
the young clusters of 0.2 Gyr to an assumed age of 12 Gyr, the LF for the 
subsample of clusters with small effective radii seems compatible with a 
Gaussian GCLF with typical parameters M$_{\rm V_0} = -7.1$ and 
$\sigma (\rm M_{\rm V_0}) = 1.3$ except for some overpopulation of the 
faint bins. These faintest bins, however, are suspected to be subject 
to the strongest depopulation through 
effects of dynamical evolution not included in our models. 
We also follow the colour evolution of the young star clusters over a Hubble 
time and compare to observations on the Milky Way and other galaxies' GC systems.

For an ongoing starburst like the one in the NGC 4038/39 system age spread 
effects among the young star cluster population may not be negligible. 
In a second step, we therefore account for age spread effects, instead of 
using a mean age for the young cluster population, and this 
drastically changes the time evolution of the LF, 
confirming Meurer's (1995) conjecture. 
We find that $-$ if age spread effects are properly accounted for $-$ the LF of 
the entire young star cluster population, and in particular that of the 
brighter subsample, after a Hubble time 
is in good agreement with the 
average Gauss-shaped LF of globular cluster systems having a turn-over at 
$\langle {\rm M_{V_0}} \rangle = -7.1$ mag and $\sigma({\rm M_{V_0}}) = 1.3$ 
mag. 
 
The age distribution shows that the brightest globular clusters from the 
interacting galaxies' original population are also observed. They make up the 
bulk of the red subpopulation with (V$-$I)$_0 > 0.95$. Their 
effective radii do not significantly differ from those of the young star 
cluster population, 
neither on average nor in their distribution.

We discuss the influence of metallicity, the effects of an inhomogeneous 
internal dust distribution, as well as the possible influence of internal $-$ 
through stellar 
mass loss $-$ and external dynamical effects on the secular evolution of the LF. 

Referring YSC luminosities to a uniform age and combining with model M/L, 
we recover the intrinsic mass 
distribution of the YSC system. It is Gaussian in shape to good approximation 
thus representing a quasi-equilibrium distribution that $-$ according to Vesperini's (1997) 
dynamical modelling for the Milky Way GC system $-$ will {\bf not} be altered 
in shape over a Hubble time of dynamical evolution, allthough a substantial 
number of clusters will be destroyed. 

We briefly compare the young star cluster population of the Antennae to the 
older one in the merger remnant NGC 7252 and point out that the intercomparison 
of young cluster populations in an age sequence of interacting and merged 
galaxies may become an interesting approach to study in detail 
the role of external dynamical effects.

\end{abstract}

\section{Introduction}

From the fact that $-$ when normalised to the stellar mass of a galaxy $-$ the 
specific globular cluster ({\bf GC}) frequency 
$T_{GC} := {N_{GC} \over {M_{\ast} / 10^9~M_{\odot}}}$ is a factor of $\sim 2$ 
higher in ellipticals than in spirals,  
Zepf \& Ashman (1993) predict that if elliptical galaxies are formed from 
one major spiral $-$ spiral merger the number of GCs formed during the merger-
induced starburst should be of the same order of magnitude as the number 
of GCs present in the progenitor galaxies.

The high burst strengths and star formation ({\bf SF}) efficiencies in massive 
gas-rich spiral $-$ spiral mergers and in IR-ultraluminous galaxies led to expect 
the formation of star clusters so tightly bound that they are able to survive 
as GCs (Fritze $-$ v. Alvensleben \& Gerhard 1994).

Fritze $-$ v. Alvensleben \& Gerhard (1994) predicted the metallicity range 
of stars and star clusters formed in massive gas-rich (i.e. late type) 
spiral$-$spiral mergers on the basis of the ISM abundances of the progenitor 
galaxies to be $\third ~{\rm Z_{\odot} \lta Z \lta Z_{\odot}}$ 
or $-0.8 \lta {\rm [Fe/H]} \lta -0.2$. 

In many interacting galaxies and merger remnants, bright blue knots have by 
now been observed (cf. e.g. Lutz 1991, Holtzman \etal 1992, Whitmore
\etal 1993, Hunter \etal 1994, O'Connell \etal 1994, 1995, 
Conti \& Vacca 1994, Borne 1996, Meurer \etal 1995). 
These bright blue knots, of course, immediately raised the 
question as to their identity: are these Young Star Clusters ({\bf YSC}) $-$ 
or, at least, some of them $-$ the 
progenitors of GCs? And, if the latter were true, how many of them are typically 
formed in a merger? How many will be able to survive 
in the tidal field of two massive interacting spirals? 
Can such a higher metallicity 
subpopulation be identified in GC systems (hereafter {\bf GCS}) 
around merger remnants and perhaps even around normal 
ellipticals? Could the metallicity distribution of a GCS give information 
about the origin of its parent galaxy (cf. Zepf \& Ashman 1993)? Or should all 
of these bright blue knots 
be open clusters/OB associations (van den Bergh 1995) most of which will 
disperse within few Gyr? The discussion of the nature of these YSCs is 
focussed on two aspects, their effective radii R$_{\rm eff}$ and their 
luminosity function. 
In mergers at distances of the Antennae or NGC 7252, effective radii as 
measured on WFPC1 images are clearly overestimated. However, it has been shown 
that for YSC systems close enough the mean effective radii do readily fall 
within the range of GC radii (Meurer \etal 1995). Our focus in this paper is 
the luminosity and colour evolution of the YSC population in the Antennae 
and, in particular, the future evolution of the YSC's LF. 

In a previous paper, we 
model the evolution of star clusters for different initial 
metallicities in terms of broad band colours and stellar metallicity indices. 
We find important colour 
differences for clusters of various metallicities, already at young ages, 
and showed that once the stellar metallicity is known, rather precise age 
dating becomes possible. 
Comparison with young star clusters in NGC 7252 (Whitmore \etal 1993), the two 
brightest of which have spectroscopy available (Schweizer \& Seitzer 1993), 
confirmed a metallicity of 
${\rm Z \sim \half Z_{\odot}}$ predicted from our global starburst modelling in 
this Sc $-$ Sc merger remnant. The mean age of the young star cluster 
population was shown to agree well with the global burst age of $\sim 1.3$ Gyr, 
and ages derived from solar metallicity models would differ by a factor 
$\sim 2$ (see Fritze $-$ v. Alvensleben \& Burkert 1995 for details). 

\medskip\noindent
Observationally, the best case by now to study the LF of YSCs are the Antennae 
with more than 700 young 
star clusters detected by Whitmore \& Schweizer (1995, hereafter {\bf WS95}), 
a number large enough to allow for a statistical analysis.
In this paper, we will examine the LF 
of the young star cluster system in the Antennae. It seems clear that not all 
bright knots in the NGC 4038/39 system with its still ongoing starburst 
will probably be GCs, in particular those 
with large effective radii R$_{\rm eff}$ might rather be open clusters or 
associations. 
Therefore, after age dating the clusters in Sect. 2., 
we subdivide Whitmore \& Schweizer's young star cluster sample into two 
subsamples containing the small knots and the more extended systems, 
respectively (Sect. 3.). In a first step, we assume a uniform age for the 
YSC population and we model the evolution of the YSCs' LF over a 
Hubble time and compare to LFs of the Milky Way's and other nearby galaxies' 
GCSs (Sect.4.). In an ongoing starburst like in the Antennae, the age spread 
among the YSCs may not be negligible (see also Meurer 1995). To 
examine the age spread effects on the LF we determe individual ages 
for all star clusters from their (V-I) colour and discuss the star clusters' 
age distribution in Sect. 5. We calculate the resulting 
individual fading for all clusters in Sect. 6. Alternative possibilities 
to subdivide the YSC sample and their 
consequences are discussed in Sect. 7. 
The of a young GCS may not only 
change by fading but also by dynamical effects 
as e.g. stellar mass loss within the cluster and/or tidal interaction of a 
cluster with the galactic potential. For GC populations in non-interacting 
galaxies, these effects were studied by Chernoff \& Weinberg (1990), their 
results are largely confirmed by the independent and more realistic approach 
of Fukushige \& Heggie (1995). In a recent paper Vesperini (1997) shows that in 
the Milky Way potential an initial log-normal mas distribution represents a 
quasi-equilibrium state that allows to preserve both its shape and parameters 
during a Hubble time of dynamical evolution, even though up to 70 \% of the 
initial cluster population get disrupted. 
In case of the Antennae, i.e. in a still uncompleted merger 
with its gravitational potential being highly variable both in space and in 
time, however, external dynamical effects seem extremely difficult 
to model. Referring YSC luminosities to a common age allows to recover the 
mass function of the YSC system when combined with model M/L. 
We discuss the possible influence 
of dynamical effects in Sect. 8. and point out the possibility to 
observationally approach these dynamical effects by intercomparing star cluster 
populations in interacting galaxies and merger remnants of various ages. 
Sect. 9. summarizes our conclusions. The spatial distribution of the YSCs $-$ 
and of their properties as derived here $-$ will be discussed in a 
forthcoming paper.

\section{Age dating of the star clusters in the Antennae}
Details of our photometric evolutionary model can be found in Fritze $-$ v. 
Alvensleben \& Burkert (1995, hereafter {\bf FB95}), where it has been used 
to age-date the young star clusters in NGC 7252. Similar to NGC 7252, though 
less advanced, NGC 4038/39 seems to be a merger of two gas-rich spirals of 
comparable mass. Though not known very accurately, the progenitor spirals of 
the Antennae may probably have been of type Sc $-$ Sc as in NGC 7252 on a similar 
kind of reasoning as in that case.  Observations of large amounts of HI within 
the body of NGC 4038/39 and along its tidal tails are from van der Hulst 
(1979), Stanford \etal (1990) report on molecular gas observations. Thus, from 
the progenitor spiral's ISM abundances a metallicity of 
$\sim \half ~{\rm Z_{\odot}}$ 
is estimated for the stars and star clusters formed during the interaction 
triggered starburst in the Antennae. 
This estimate would not change by much, if e.g. one of the progenitor spirals 
were of type Sb or Sd. As long as no spectroscopic abundance determination is 
available for young star clusters in NGC 4038/39, 
we will have to rely on this rough metallicity estimate. While a certain 
metallicity spread among young star clusters formed in the burst cannot be 
excluded, in a first step 
we will $-$ for lack of better knowledge $-$ assume that all young clusters 
have this same metallicity of $\half ~{\rm Z_{\odot}}$ and we will derive 
a mean age of the YSC population from their mean ${\rm (V-I)}$ colour. 

In a second step, we release the simplifying assumption that all YSCs have the 
same age and we will 
derive ages for individual clusters from their ${\rm (V-I)}$ to discuss the effect of 
an age spread among the very young star clusters. 

For a mean dereddened $\langle {\rm (V-I)_0} \rangle_{({\rm all ~ clusters})} \sim 0.5$ 
(cf. WS95), our models give a mean cluster age of $\sim 2 \cdot 10^8$ yr 
(cf. FB95). 
If the metallicity were as low (as high) as $1 \cdot 10^{-3}$ (as $2 \cdot 
{\rm Z_{\odot}}$) the clusters would be ascribed ages of $\sim 4 \cdot 10^8$ yr 
($\sim 1 \cdot 10^8$ 
yr). These mean ages of the YSC population seem quite compatible with Barnes' 
(1988) dynamical time of $2\cdot 10^8$ yr since the last (=first) pericenter. 

${\rm (U-V)}$ colours are available for only 48 YSCs. Their mean dereddened 
$\langle {\rm (V-I)_0} \rangle \sim -1.0$ would lead to a mean age of 
$\sim 1 \cdot 10^7$ yr for ${\rm Z \sim \half Z_{\odot}}$. 
This younger age is not 
in conflict with our $\langle {\rm age} \rangle \sim 2 \cdot 10^8$ yr from 
${\rm (V-I)}$ 
since only the very brightest YSC in U are detected which $-$ of course $-$ are 
expected to be the bluest and youngest.

Both nuclei of NGC 4038 and 4039 are sites of ongoing or recent 
strongly enhanced star formation 
(e.g. Rubin \etal 1970, Keel \etal 1985) and contain sufficient reservoirs of 
molecular gas (Stanford \etal 1990) to sustain their starbursts for a while. A 
typical burst duration in this kind of gas-rich spiral $-$ spiral merger is of 
the order of $\sim 4 \cdot 10^8$ yr (Fritze $-$ v. Alvensleben \& Gerhard 1994, 
Bernl\"ohr 1990, Carico \etal 1990).

\section{Subdivision of the Antennae's YSC sample with respect to R$_{\rm eff}$} 
In the course of mass loss through any mechanism whatsoever during secular 
evolution, the effective radius of a star cluster can only grow (cf. Sect. 8). 
From the observed range 
of effective radii R$_{\rm eff} = 0 \dots 50$ pc (we use H$_0 = 75$ throughout) 
it seems clear that not all bright knots in the Antennae may be GCs, the more 
extended ones having a higher probability of being open clusters or associations. 
Here, we are facing a significant difference to the case of NGC 7252 where from 
the larger mean age of $\sim 1.3$ Gyr alone, most objects can be expected to be 
GCs. And indeed, the mean effective radius of the NGC 7252 clusters 
is $\sim 7$ pc, significantly smaller than the $\langle {\rm R_{eff}} \rangle 
= 13$ pc 
of all the Antennae's clusters, despite the considerably larger distance to 
NGC 7252.

Meurer (1995) argues that $-$ due to severe crowding on a bright and spatially 
variable galaxy background $-$ YSC effective radii in 
distant galaxies may be strongly overestimated. Effective radii are generally 
determined from the luminosity difference in a small and a larger aperture 
centered on a YSC, where the large aperture in some cases might be contaminated 
by light from neighbouring star clusters. The observed clustering of YSCs $-$ 
typically a dozen within one giant HII region (WS95) $-$ tends to increase 
this overestimation of effective radii. Meurer \etal (1995) estimate the distance 
out to which this two aperture method should be expected to yield reliable 
R$_{\rm eff}$ to be 9 Mpc and emphasise that the mean R$_{\rm eff}$ of YSCs 
in all three starburst galaxies observed to date within this distance is indeed 
$\sim 1.3$ pc, i.e. even smaller than the median R$_{\rm eff} \sim 3$ pc of 
Galactic GCs as given by Djorgovski \& Meylan (1994). 

Tidal radii $R_T$ or core radii $R_C$ could not be determined for The 
Antennae's clusters, so no information is available about their concentration 
parameters \\
${\rm c := Log R_T / R_C}$ (${\rm = log R_T/R_{eff}}$ before core 
collapse) \\
which are crucial for the question of survival or destruction (Chernoff \& 
Weinberg 1990). Thus, we are left with effective radii as the only 
discriminating quantity between probable GCs and suspected open clusters 
(but see also Sects. 4 \& 5). 

In the following, we devide WS95's original YSC sample into two 
subsamples of clusters with R$_{\rm eff} \leq 10$ pc, which probably will 
contain young GCs, 
and of clusters with R$_{\rm eff} >10$ pc, which may contain 
open clusters or OB associations, but possibly young GCs, 
too. Galactic GCs have effective radii in the range of 1 $-$ 25 pc 
(Djorgovski \& Meylan 1994). We chose 
a delimiting R$_{\rm eff}$ of 10 pc in order not to have too low a number of 
objects in the small  R$_{\rm eff}$ subsample. 

Table 1 compares the mean properties of the two subsamples. 
All throughout this paper, numbers after the $\pm$ sign are standard errors. 
It is seen that 
the subsample with R$_{\rm eff} \leq 10$ pc has $\langle {\rm R_{eff}} \rangle 
= 6.9 \pm 2.2$ pc as compared to the mean effective radius of Galactic GCs 
of $\sim 3$ pc, while the subsample with R$_{\rm eff} >10$ pc has 
$\langle {\rm R_{eff}} \rangle = 16.6 \pm 5.5$ pc. 

Furthermore, Table 1 shows that clusters with R$_{\rm eff} \leq 10$ pc, as 
compared to clusters with larger R$_{\rm eff}$, have slightly 
larger mean galactocentric distances R$_{\rm gc}$, a marginally higher V $-$ 
luminosity (by 0.16 mag), are redder in ${\rm (V-I)}$ by 
0.02 and in ${\rm (U-V)}$ by 0.08 mag with a smaller scatter in their colours. 
It is worth noting that if the observed colour 
difference is interpreted in terms of an age difference, our evolutionary 
model for ${\rm Z = 0.01}$ indeed indicates a 
luminosity difference compatible with the one observed. 
WS95 give an average correction for internal reddening of the YSCs of 
$\Delta {\rm (V-I)} = 0.3$ mag. When compared to our ${\rm Z =\half Z_{\odot}}$ 
models, the 
dereddened $\langle {\rm V-I} \rangle_0 = 0.45 \pm 0.44$ of the YSCs with 
R$_{\rm eff} \leq 10$ pc corresponds to a mean age of $2 \cdot 10^8$ yr with an 
age dispersion ranging from $1 \cdot 10^7$ to $2 \cdot 10^9$ yr, while with  
$\langle {\rm V-I} \rangle_0 = 0.43 \pm 0.49$ the clusters with R$_{\rm eff} > 
10$ pc have a mean age of $1.6 \cdot 10^8$ yr with an even larger dispersion 
ranging from $6 \cdot 10^6$ to $2.5 \cdot 10^9$ yr. 

With this mean age and the 
assumed metallicity the YSCs with R$_{\rm eff} \leq 10$ pc will redden until 
an age of 12 Gyr (15 Gyr) to a $\langle {\rm B-V} \rangle_0 \sim 0.85$ ($\sim 0.88$, 
respectively). This is significantly redder than the observed 
$\langle {\rm B-V} \rangle_0$ of the Milky Way, Andromeda, LMC or SMC GCs which 
are in the range 0.67 $-$ 0.74 (Harris \& Racine 1979). The red $\langle {\rm B-V} 
\rangle_0$ is due to the higher 
mean metallicity of these secondary generation clusters. 
GC systems in ellipticals have a mean metallicity typically higher by 0.5 dex 
than that of spiral galaxy GC systems and are therefore expected to have 
$\langle {\rm B-V} \rangle_0 \sim 0.93$ (Ashman \etal 1995). The reddest GC system 
known is that of the Hydra cD NGC 3311 with $\langle {\rm [Fe/H]} \rangle = -0.31$ dex 
(Secker \etal 1995). Ashman \etal's analysis using Worthey's (1994) models as 
well as our own models (FB95) suggest a $\langle {\rm B-V} \rangle_0 \sim 1.0$ for 
this extreme GC system, while for the metallicity we assume for the young 
Antennae clusters a $\langle {\rm B-V} \rangle_0 \sim 0.9$ is predicted by Ashman 
\etal, close to the 15 Gyr value we obtain. 

{\bf To conclude}, 
all the differences between the two YSC subsamples do not prove but are 
consistent with a scenario of a global starburst contracting in time with 
the YSCs now observed with R$_{\rm eff} \leq 10$ pc  formed 
on average in a 
slightly earlier and spatially somewhat more
extended stage of the starburst with an age spread slightly smaller than that of 
the clusters with R$_{\rm eff} > 
10$ pc that might still be forming now, at a somewhat later phase in a  
starburst region that already has contracted towards the center and/or with a 
star formation efficiency that has decreased at large 
radii. 

\begin{table}
\caption{Comparison of young star cluster subsamples.}
\halign{\hfil#\hfil&\hfil#\hfil&\hfil#\hfil \cr
\noalign{\medskip\hrule\medskip} \cr
 & R$_{\rm eff} \leq 10$ pc & R$_{\rm eff} >10$ pc \cr
\noalign{\medskip\hrule\medskip} \cr
 N$_{\rm obj}$  & 242  & 472  \cr
 $\langle {\rm R_{gc}} \rangle$  & $3.65 \pm 1.76$ kpc & $3.49 \pm 1.58$ kpc \cr
 $\langle {\rm V} \rangle$ & $~21.57 \pm 1.17$ mag ~ & $~21.73 \pm 1.00$ mag ~ \cr
 $~\langle {\rm V-I}  \rangle ~$ & $~0.75 \pm 0.44$  & $~0.73 \pm 0.49$ \cr
 $\langle {\rm U-V}  \rangle$  & $-0.66 \pm 0.23$  & $-0.74 \pm 0.26$ \cr
 $\langle {\rm R_{eff}} \rangle$ & $6.92 \pm 2.15$ pc & $16.60 \pm 5.46$ pc \cr
\noalign{\medskip\hrule\medskip}}
\end{table}

\section{Evolution of the YSCs' Luminosity Functions over a Hubble time}

In Fig. 1., we present the LFs of the star cluster subsamples with R$_{\rm eff} 
\leq 10$ pc (Fig. 1a) and with R$_{\rm eff} >10$ pc (Fig. 1b). We have 
transformed apparent to absolute luminosities by using a distance 
modulus of 31.42 corresponding to a distance of 19.2 Mpc (H$_0 = 75$) to 
NGC 4038/39. 
 
\begin{figure} 
\includegraphics[width=\columnwidth]{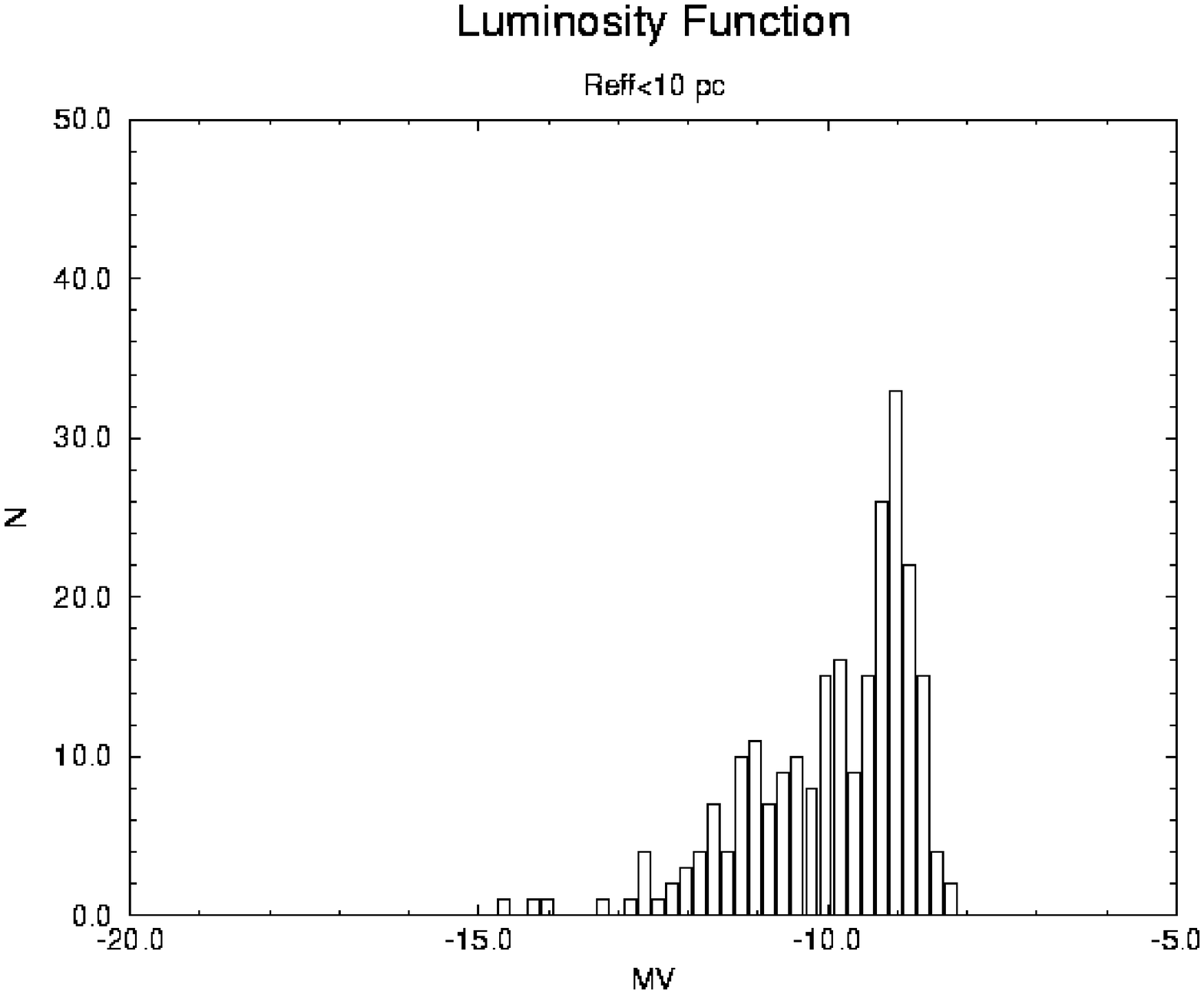}
\includegraphics[width=\columnwidth]{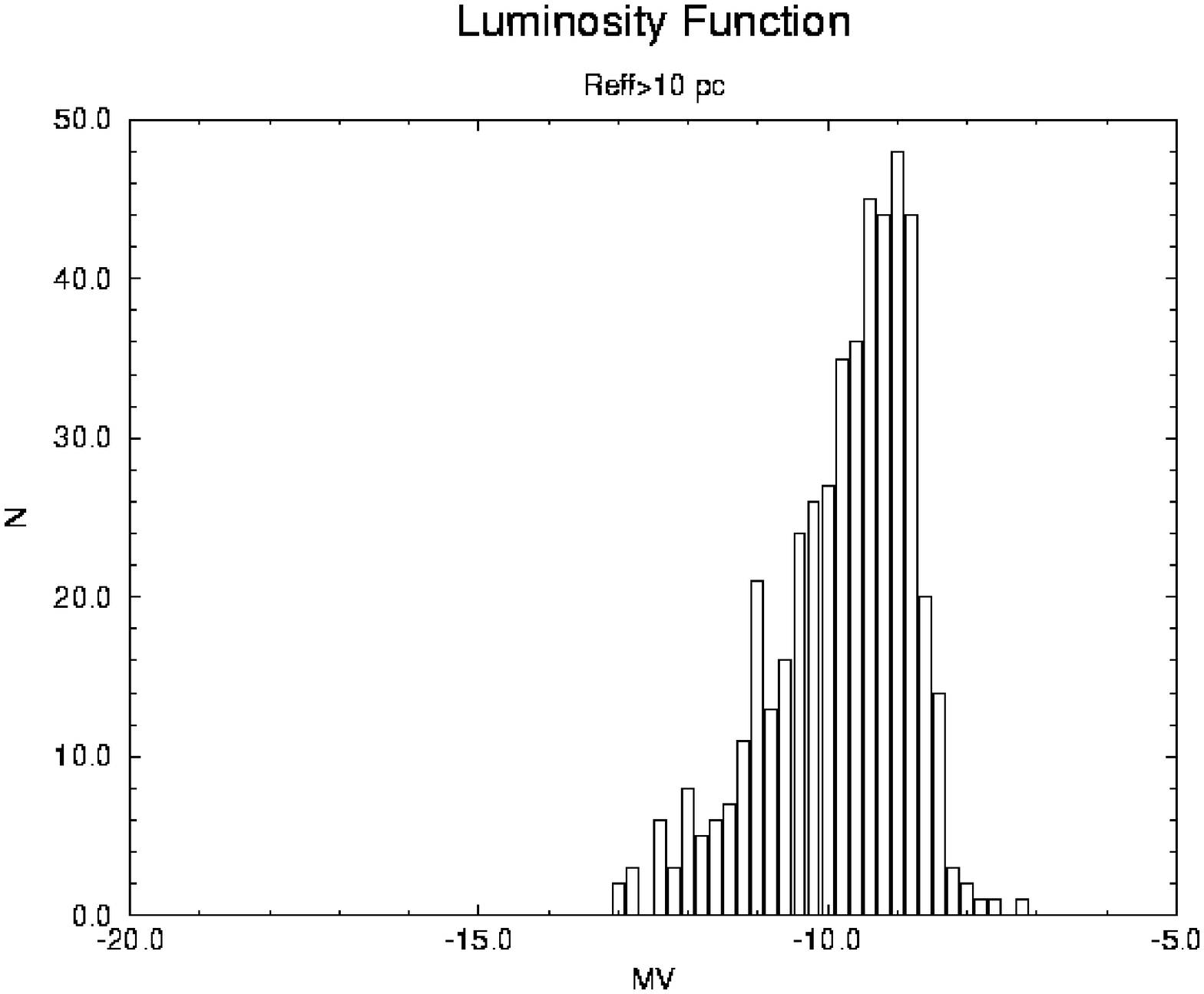}
 \caption{LFs for star clusters with R$_{\rm eff} \leq 10$ pc {\bf (1a)} and 
 for star clusters with  
R$_{\rm eff} >10$ pc {\bf (1b)} in the Antennae as observed by WS95.} 
\end{figure}

\begin{figure} 
\includegraphics[width=\columnwidth]{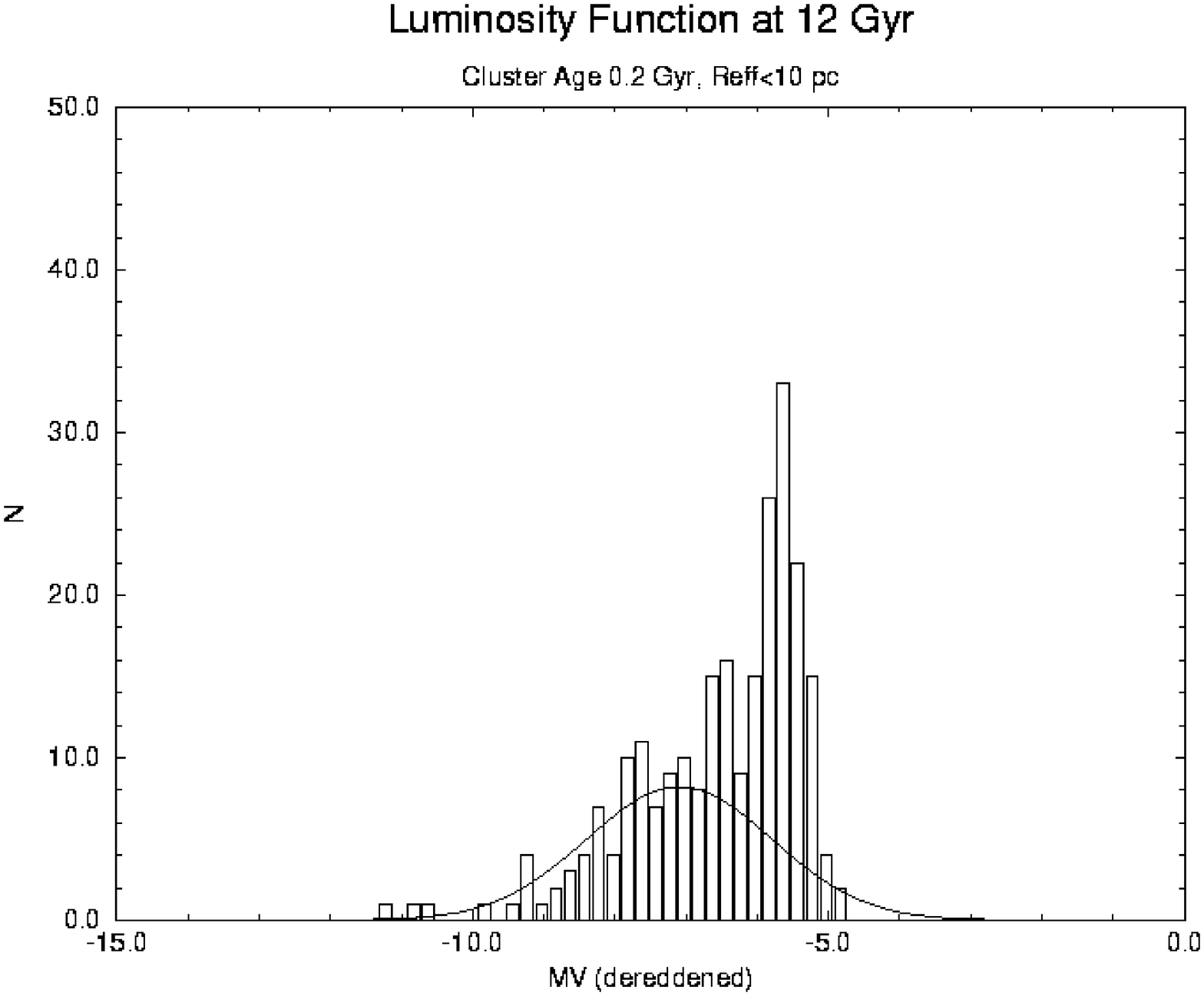}
\includegraphics[width=\columnwidth]{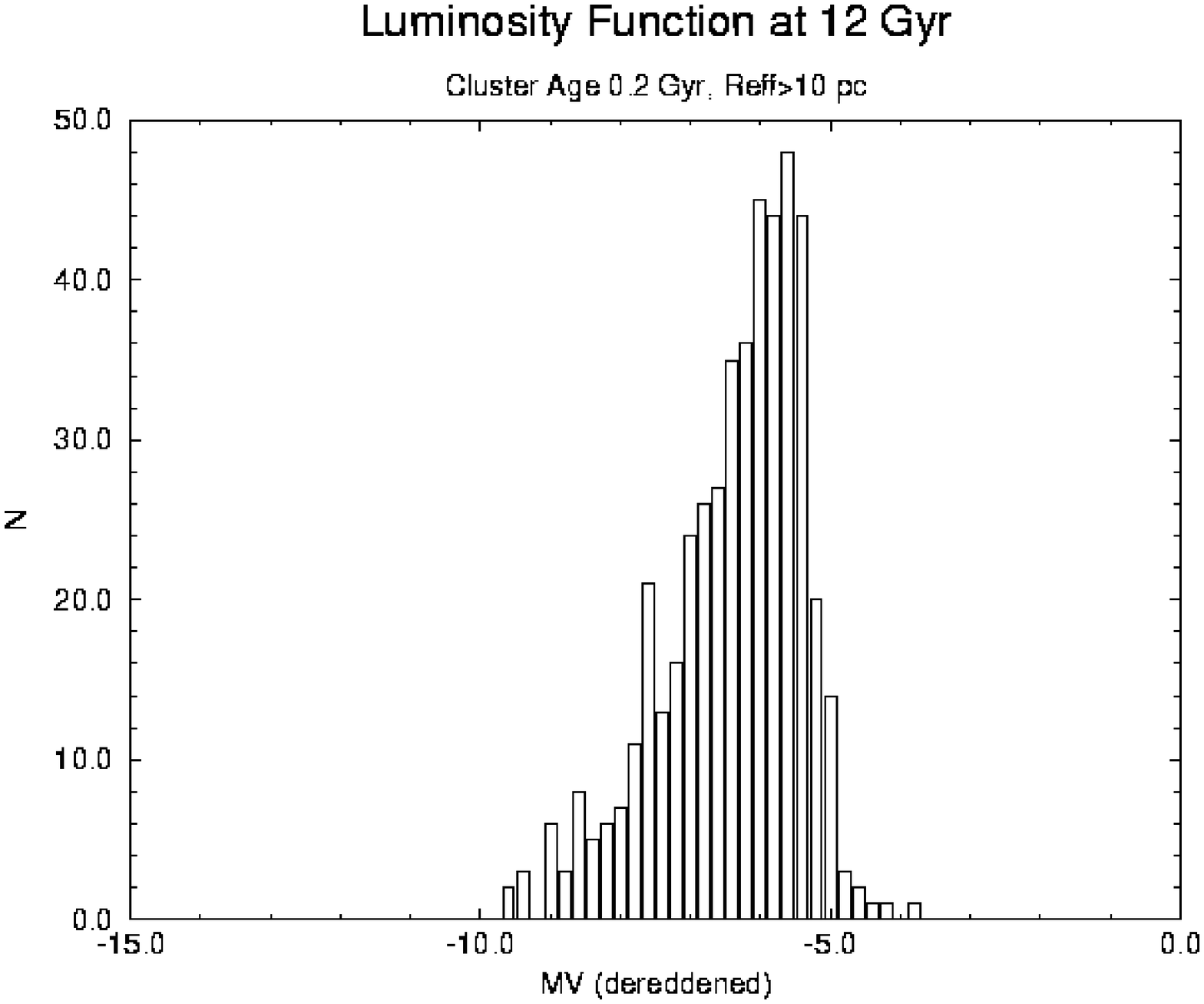}
 \caption{Dereddened LFs for star clusters with R$_{\rm eff} \leq 10$ pc 
 {\bf (2a)} and 
 for star clusters with R$_{\rm eff} >10$ pc {\bf (2b)} at a cluster age of 
 12 Gyr as given by our models. Superimposed on Fig.2a. is a Gaussian 
with $\langle {\rm M_{V_0}} \rangle = -7.1$ mag and $\sigma_{\rm M_V} =1.3$ mag, 
scaled to the total number of YSCs with R$_{\rm eff} \leq 10$ pc in 
the Antennae.} 
\end{figure}

Assuming an initial metallicity of $\half ~{\rm Z_{\odot}}$ and an average age of the 
young star clusters of 0.2 Gyr as derived in Sects. 1. and 2., our models give 
the (purely photometric) luminosity evolution in various passbands and allow 
to calculate the LF at any time from the presently observed one. 
In Fig. 2., we show the LF at a time of 12 Gyr for the subsamples of clusters 
with R$_{\rm eff} \leq 10$ pc (Fig. 2a), and for clusters with  R$_{\rm eff} > 
10$ pc (Fig. 2b). Dereddened luminosities are obtained by applying a constant 
internal dust extinction correction of ${\rm A_V} = 0.5$ mag (cf. WS95) for all YSCs. 
Going to a later time of 
15 Gyr would simply shift the LF to the fainter side by 0.3 mag in Figs 2a, b. 
For comparison, we 
also depict in Fig. 2a. a Gaussian type LF with $\langle {\rm M_{V_0}} \rangle 
= -7.1$ mag and 
$\sigma({\rm M_V}) = 1.3$, the average values given by Harris (1991) for the GCSs 
of 16 galaxies, normalised to twice the number of clusters with  
R$_{\rm eff} \leq 10$ pc and M$_{\rm V_0} \leq -7.1$ mag. 
It is seen that the LFs for the subsamples of clusters with small and large 
effective radii look intriguingly different. The LF for 
YSCs with 
R$_{\rm eff} > 10$ pc indeed looks exponential like the open cluster LF in the 
Milky Way or the Magellanic Clouds. 
The LF of YSCs with R$_{\rm eff} \leq 10$ pc does not show a turnover up to 
the completeness limit, which has evolved to ${\rm M_{V_0}} = -5.7$ mag 
at 12 Gyr. Yet one may argue that to some degree it 
resembles the Milky Way and other galaxies' GC LF with a strong 
overpopulation of the faint magnitude bins. 
It is, however, the faintest bins that may be expected to be most severely 
depopulated in the course of dynamical evolution (cf. Sect. 8.). 
Moreover, in an ongoing starburst as in the NGC 4038/39 system, our simplifying 
assumption that the YSCs are coeval clearly is a poor approximation and 
age spread 
effects will redistribute the final star clusters' luminosities as shown in 
Sect. 6. A Kolmogorov-Smirnov ({\bf KS}) test shows the probability that 
the LFs of both subsamples of clusters are obtained from the same parent 
population is 20 \% if all clusters are considered and 4 \% for clusters 
brighter than the completeness limit ${\rm M_{V_0}} = -9.6$ mag. 

The observed crowding of YSCs may raise the suspicion that some of the apparent 
large ${\rm R_{eff}}$ clusters are in fact blended pairs of small 
${\rm R_{eff}}$ clusters. To test for a possible contamination of the large 
${\rm R_{eff}}$ sample by unresolved pairs we extrapolate the LF of the small 
${\rm R_{eff}}$ clusters to fainter magnitudes (to M$_{\rm V_0} = -7.6$ mag), 
randomly draw pairs from this extrapolated LF and compare the LF of pairs to the 
LF of the large ${\rm R_{eff}}$ clusters. The LF of pairs contains a significant 
number of clusters brighter than the brightest clusters in the large 
${\rm R_{eff}}$ subsample and a larger number of bright clusters. 
A KS $-$ test shows the contamination 
of the large ${\rm R_{eff}}$ sample by unresolved pairs is $\lta 10$ \% at levels 
brighter than the completeness limit.

If instead of subdividing the Antennae's YSC 
sample at R$_{\rm eff} = 10$ pc we  divide at R$_{\rm eff} = 5$ pc 
the statistics becomes very poor for the YSCs with R$_{\rm eff} \leq 5$ pc but 
no qualitative changes of the LFs are indicated.

{\bf We conclude} from Fig. 2 that the LFs evolved to an age of 12 Gyr 
of YSCs with R$_{\rm eff} > 10$ pc and R$_{\rm eff} \leq 10$ pc are significantly 
different. The similarity of the small R$_{\rm eff}$ clusters' LF 
with GC systems' LFs does question 
the use of the LF as an argument against them being young GCs, 
as was done by van den Bergh (1995) for the entire sample using 
solar metallicity models from Bruzual \& Charlot (1993). Unresolved close pairs 
of YSCs do not contribute to the large R$_{\rm eff}$ cluster sample by more than 10 \%.

\begin{figure} 
\includegraphics[width=\columnwidth]{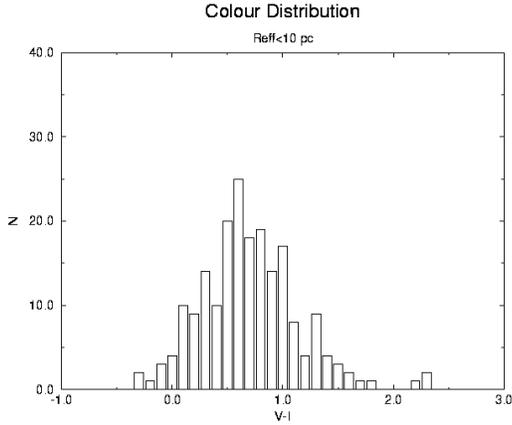}
\includegraphics[width=\columnwidth]{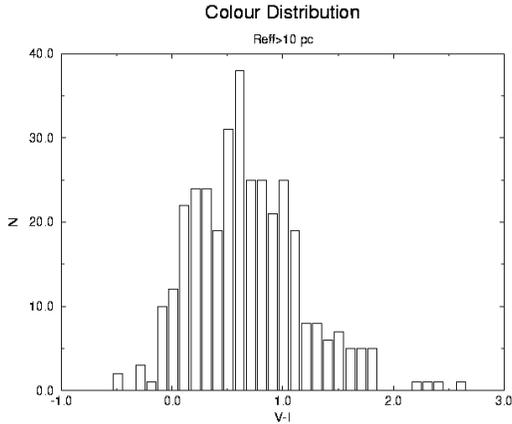}
 \caption{${\rm V-I}$ colour distribution of star clusters with R$_{\rm eff} 
 \leq 10$ pc {\bf (3a)} and of star clusters with  R$_{\rm eff} >10$ pc 
 {\bf (3b)} at the present time as derived from WS95's observations.}
 \end{figure}

Fig. 3 presents the colour distribution of the YSCs with R$_{\rm eff} \leq 
 10$ pc (Fig. 3a) and of clusters with  R$_{\rm eff} >10$ pc 
(Fig. 3b) as it is presently observed (WS95). For the more extended clusters the 
colour distribution is broader than for the small R$_{\rm eff}$ subsample and 
slightly shifted to the blue. A KS-test shows that the colour distributions 
are drawn from different populations with 85 \% probability for clusters 
brighter than the completeness limit. 

\begin{figure} 
\includegraphics[width=\columnwidth]{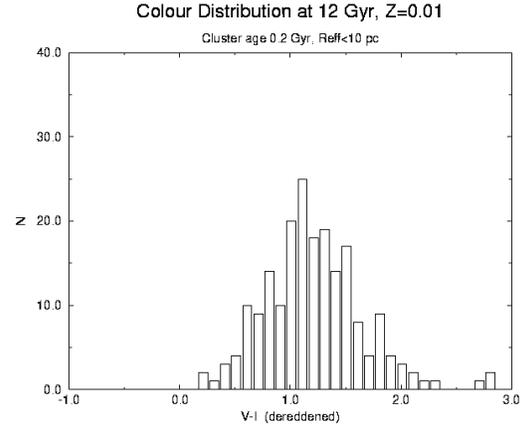}
\includegraphics[width=\columnwidth]{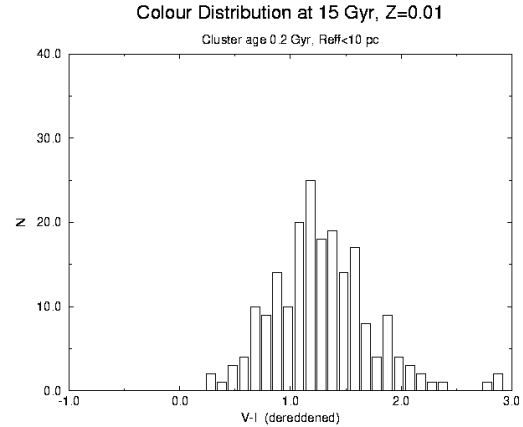}
\caption{Dereddened ${\rm (V-I)_0}$ colour distribution of star clusters 
 with R$_{\rm eff} \leq 
 10$ pc at a time  of 12 Gyr {\bf (4a)} and at a time of 15 Gyr {\bf (4b)} as 
 calculated from our models.}
 \end{figure}

Fig. 4 shows the dereddened colour distribution of the small R$_{\rm eff}$ 
subsample after 12 (Fig. 4a) 
and 15 Gyr (Fig. 4b) of undisturbed evolution. With $\langle {\rm V-I} 
\rangle_0 =1.15$ and $\langle {\rm V-I} \rangle_0 =1.23$ at 12 Gyr and 15 Gyr, 
respectively,  
the mean colour of YSCs in the Antennae, when only passively aged for our 
assumed metallicity of Z$= 0.01$ and a common age of the YSCs of 0.2 Gyr, 
is very close to the mean $\langle {\rm V-I} 
\rangle =1.20$ of the Milky Way halo GC system. 

We caution, however, that these colours distributions simply were obtained 
by shifting the observed colour distribution (Fig. 3a) by the amount of 
reddening given by our evolutionary models during aging of the YSCs. 
Age spread effects that change the LF will also affect 
the colour distribution as shown in the next Sect. We expect the red clusters 
to be older than average and therefore to redden less during further evolution, 
while blue clusters may tend to redden more. Thus, we expect age spread 
effects to reduce the width of the colour distribution over a Hubble time.

\section{Age distribution of star clusters in the Antennae: Young star 
clusters and old globular clusters}

In an ongoing starburst as in the Antennae, the age spread among YSCs can be 
comparable to their ages, and age spread effects  may be expected to 
significantly affect the time evolution of the LF (see also Meurer 1995). 

We therefore, in a second step,  
derive individual ages for the YSCs from their individual ${\rm (V-I)}_0$, dereddened 
using a 
common internal dust correction $\Delta {\rm (V-I)} = 0.3$ mag as given by WS95 for 
all clusters, assuming that all YSCs have the same metallicity Z$ = \half 
{\rm Z_{\odot}}$.

\begin{figure} 
\includegraphics[width=\columnwidth]{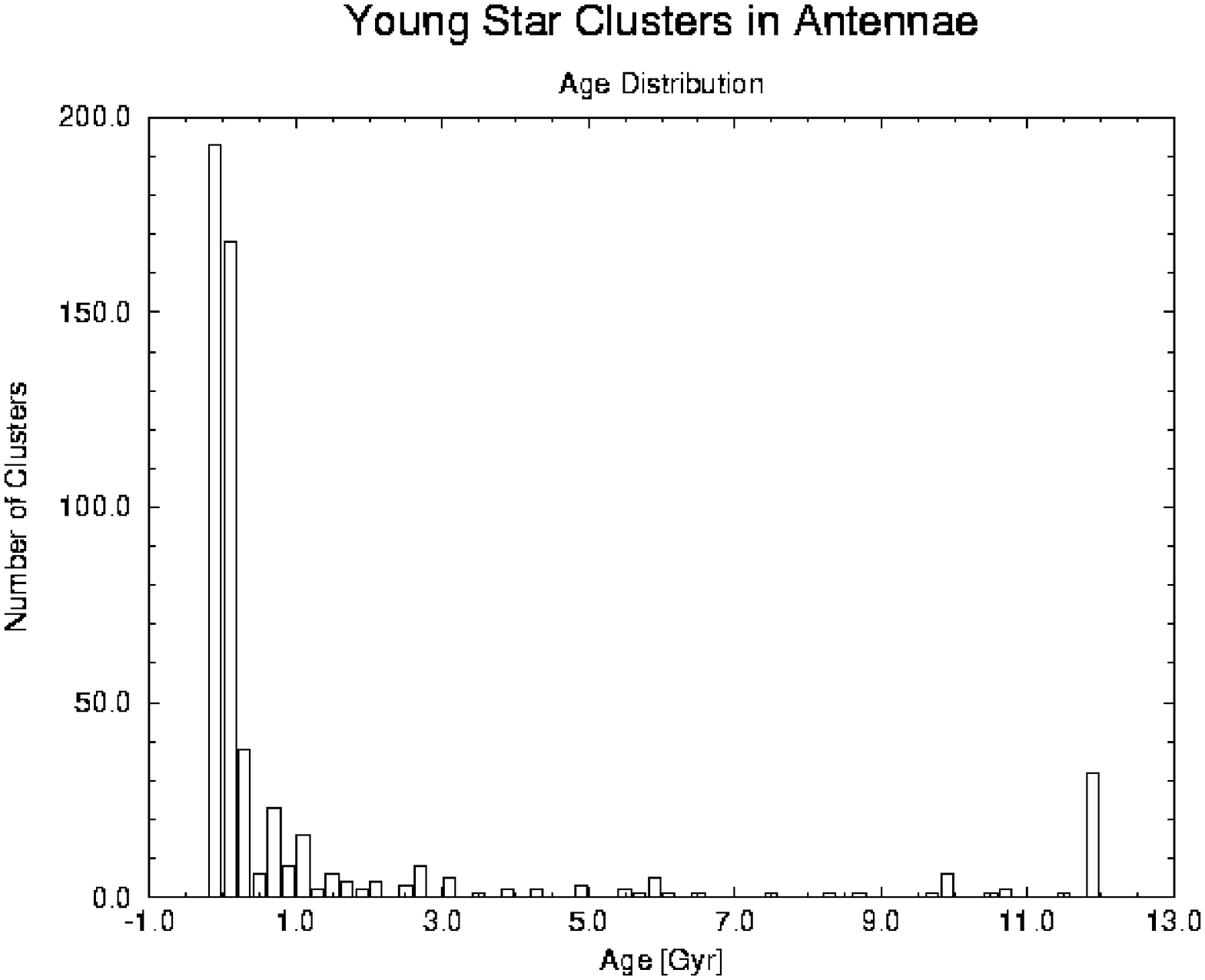}
\includegraphics[width=\columnwidth]{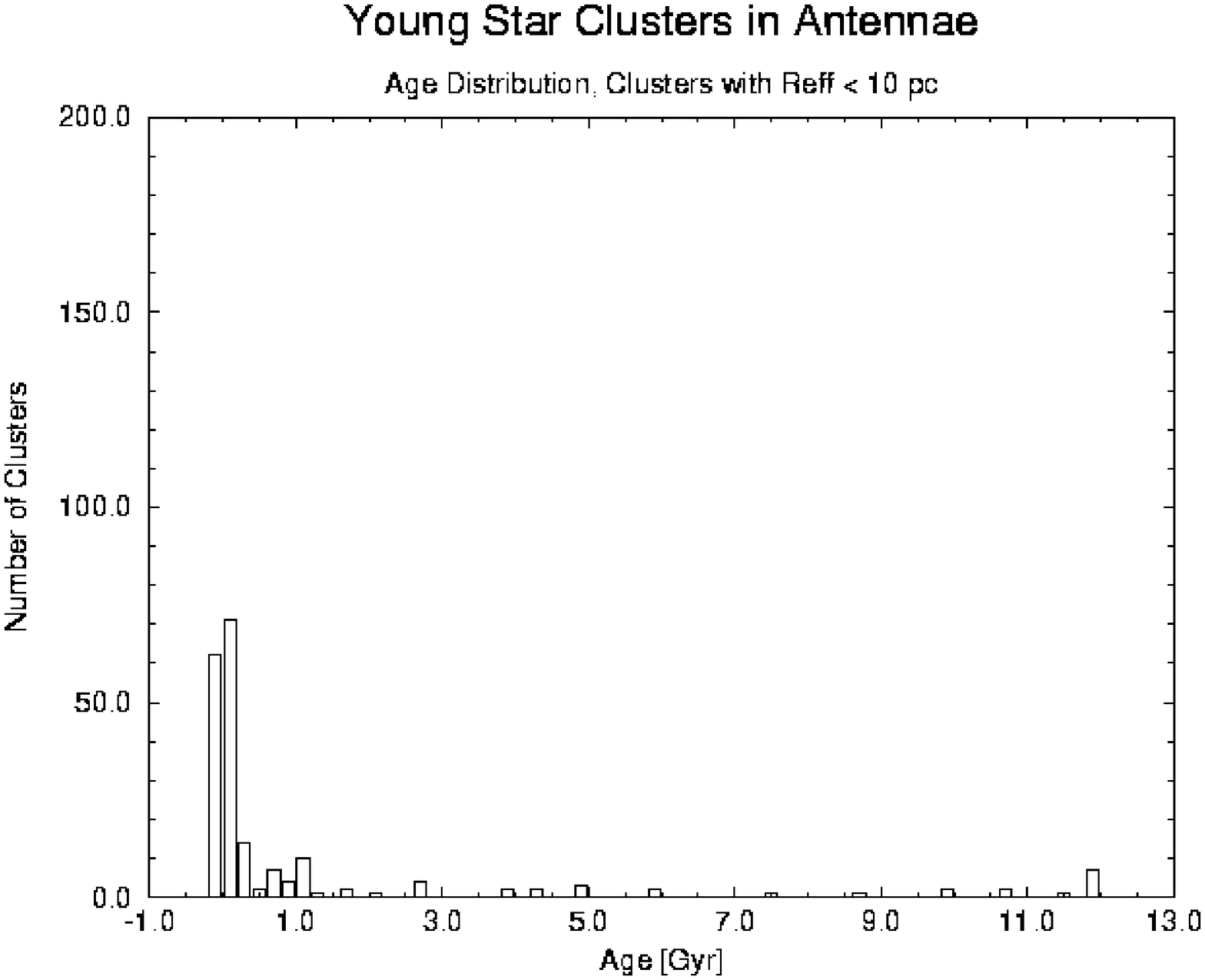}
\includegraphics[width=\columnwidth]{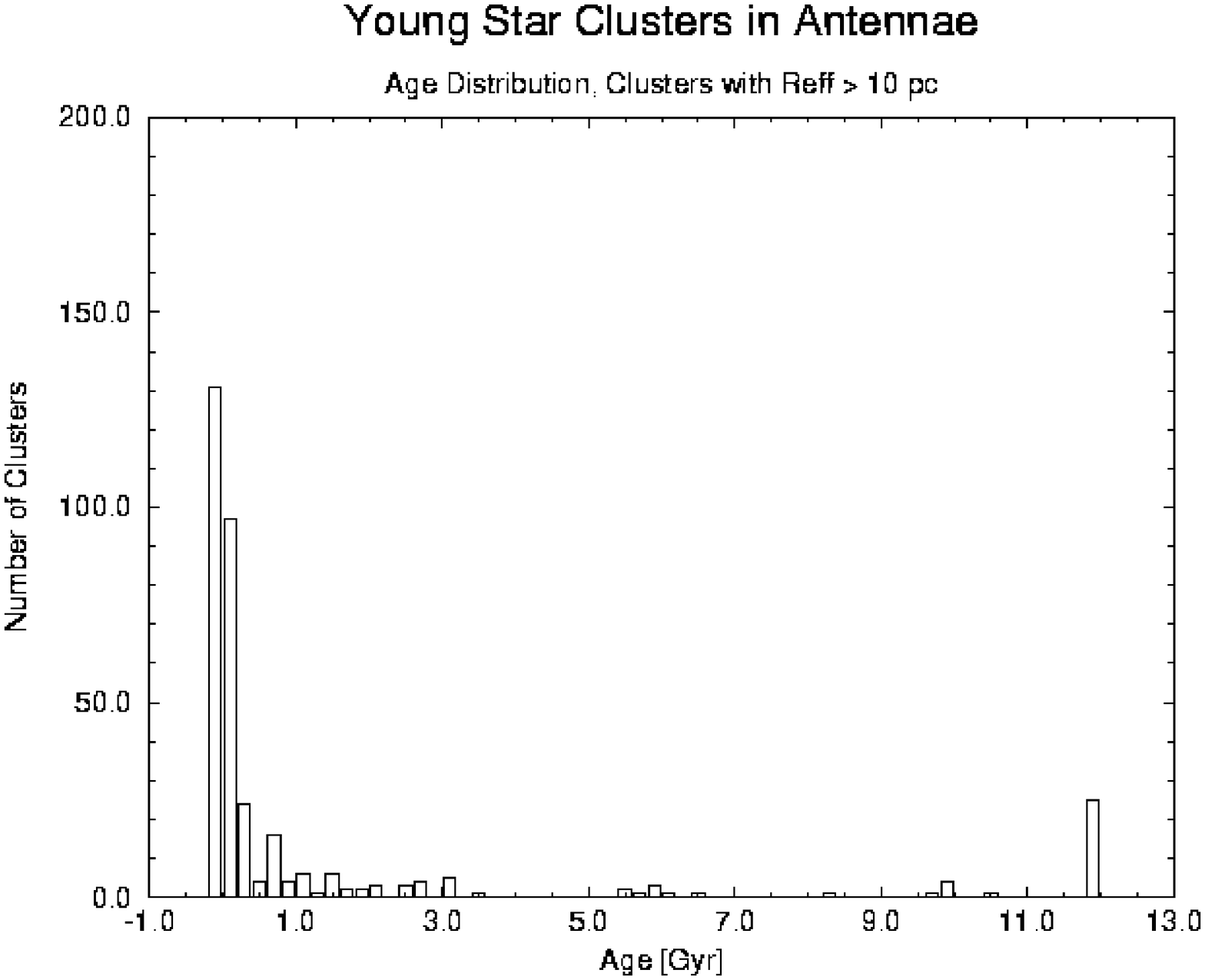}
 \caption{Age distribution of star clusters in the Antennae. {\bf (5a)} 
 all clusters, {\bf (5b)} clusters with R$_{\rm eff} \leq 
 10$ pc, {\bf (5c)} clusters with  R$_{\rm eff} >10$ pc.}
 \end{figure}

In Fig. 5a, we present the age distribution of all star clusters in the Antennae. 
It shows two strictly distinct peaks, a very strong one at ages 
$(0 - 4)\cdot 10^8$ 
yr and a smaller one at $\sim 12$ Gyr. 
Out of the 714 clusters of WS95, 164 do not allow for age determination because 
they lack I $-$ band observations. Of the remaining 550 clusters, 399 have ages 
$(0 - 4)\cdot 10^8$ yr, 32 are as old as 12 Gyr, provided they do not suffer 
from much larger than average extinction. 119 are interloopers with 
$4 \cdot 10^8$ yr $<$ age $< 12$ Gyr, most of them with ages below 1 Gyr. 

If the 37
clusters with apparent ages $3 {\rm ~Gyr} < {\rm age} < 12$ Gyr had a lower 
metallicity of 
${\rm Z=1\cdot10^{-3}}$ or ${\rm Z=1\cdot10^{-4}}$, they might well have ages of 12 or 
15 Gyr, respectively. On the other hand, if the 82 clusters with ages in 
the range 
$4\cdot10^8$ to $<3\cdot10^9$ yr were in an environment with a higher than 
average dust reddening or if they had a metallicity ${\rm Z > 0.01}$, their ages 
could easily be reduced to $\lta 4\cdot 10^8$ yr. Thus, we believe that there is no 
convincing evidence for the existence of intermediate age clusters, rather we 
expect a non-homogeneous internal dust distribution and an intrinsic scatter in 
metallicity comparable to the one observed for every GC system to be responsible 
for the apparent interloopers. 
The 32 old clusters from the 12 Gyr peak in Fig. 5a and the 37 interloopers 
with ages $\geq 3$ Gyr for which we argue that their ages may be underestimated 
must be part of the original GC 
population of the interacting spirals NGC 4038 and 4039. 
53 of them are 
brighter than the completeness limit of ${\rm M_V} = -9.1$ mag. 
Comparing to the Milky Way and M31 GCLFs we find that 10/131 GCs (Milky Way) 
and 27/200 GCs (M31) are brighter than 
the completeness limit. By analogy one would expect NGC 4038 and NGC 4039, 
together, to have had of the order of 20 $-$ 50 GCs brighter than -9.1 mag. 

The number ratio of young to old clusters is $\sim 12$, while 
including the interloopers the number ratio 
of probably young to probably old clusters drops to $\sim 7$. 

One might doubt that the 69 objects we tentatively identify as old GCs from the 
Antennae system's progenitor spirals might in fact be extremely reddened YSCs. 
They have ${\rm (V-I)}$ colours in the range 1.3 $-$ 2.7 mag. If they were YSCs 
with their red colours entirely due to larger than average internal extinction, 
their V $-$ magnitudes should be affected by ${\rm A_V} =$ 1.3 $-$ 3.8 mag which 
is much larger than the observed luminosity difference between the red and 
blue star cluster subsamples (cf. Sect. 7.1). Moreover, visual inspection of the 
projected distribution of what we chose to call old GCs shows that they are not 
as strongly clustered as are the blue YSCs, nor do they trace the internal 
tidal structure of NGC 4038/39 as do the blue clusters. While a few of the 
very red clusters lie close 
to very blue ones, their overall distribution looks much more spherically 
symmetric than that of the blue clusters.

WFPC2 imaging is expected to go $\gta 2$ mag deeper for the Antennae (WS95) 
and thus should reach or come close to the turn-over of the orinal GCSs' LF. 
This would then allow for a reasonable statistical analysis of the spatial 
distribution of the red clusters, for a 
reliable estimate of the progenitor spirals' 
total number of GCs, and for a detailed comparison of the old GCS and the YSC 
system. Until then, we cannot exclude the possibility that a small fraction 
of the red clusters may not belong to the original GC population but rather be 
exceptionally reddened bright YSCs. 

38 out of the 48 YSCs with ${\rm (U-V)}$ colour available allow for age dating while 
10 have ${\rm (U-V)}$ bluer than our $\half ~{\rm Z_{\odot}}$ model ever reaches. 
Either their extremely blue ${\rm (U-V)}$ is influenced by gaseous emission, or their 
metallicity is particularly low, or $-$ perhaps most plausibly $-$ they are 
affected by less than average reddening inside the Antennae. For the 38 YSCs with 
${\rm (U-V)_0} \geq -1.15$, i.e. within the range of our $\half ~{\rm Z_{\odot}}$ model, 
ages derived from ${\rm (U-V)_0}$ agree with those derived from ${\rm (V-I)_0}$ to within 
$\leq 1 \cdot 10^7$ yr. 

Figs. 5b and 5c present the age distributions of compact and extended clusters, 
respectively. They are clearly different  as confirmed by a KS-test, the 
probability that they are drawn from the same parent population is 
$4 \cdot 10^{-7}$. It is interesting to note that the number ratio of YSCs to 
old GCs is larger among the compact cluster subsample than among the extended 
clusters: ${\rm N_{YSC}/N_{GC} \sim 21}$ for clusters with 
${\rm R_{eff} \leq 10}$ pc as compared to ${\rm N_{YSC}/N_{GC} \sim 10}$ 
for clusters with ${\rm R_{eff} > 10}$ pc. If we restrict the age analysis to 
clusters brighter than the completeness limit, we preferentially lose old GCs 
because they tend to be fainter than the bulk of the YSC population. 
This increases the ratio ${\rm N_{YSC}/N_{GC}}$ by about a factor of 2. 
Nevertheless, the number ratio of young to old clusters remains larger by 
almost a factor 2 among the small ${\rm R_{eff}}$ clusters  than among the 
large ${\rm R_{eff}}$ clusters, contrary to what would be expected  if the bulk of 
the YSCs formed in the merger were open clusters instead of proto-globulars. 
Restriction to clusters brighter than ${\rm M_{V_0} = -9.6}$ mag drastically 
reduces the relative number of interloopers in the age distribution. This supports 
our argument in favour of 2 distict episodes of cluster formation separated by 
a time span of more than 10 Gyr.

{\bf In summary}, it turns out that the bright star cluster population 
detected with WFPC1 in the Antennae contains an important fraction of 
$\sim 12$ Gyr old objects (69 out of 550 clusters for which age dating is possible) 
most of which seem to be part of the bright end of the original spirals' GC 
population. The number ratio of YSCs to old GCs is larger by a factor $\sim 2$ 
among the small ${\rm R_{eff}}$ subsample than among the extended clusters.

\section{Age spread effects on the LF}

Meurer (1995) pointed out the possible importance of age spread effects on 
the future luminosity evolution of YSCs. 
Faint clusters, on 
average, will tend to be older than bright ones. As a consequence, 
they will fade less during further evolution, thus migrating from 
fainter to brighter bins in the LF. Clusters brighter than average, by analogy, 
tend to evolve to fainter bins of the LF. 

So now, from the individual cluster ages we derived in Sect. 5, we 
calculate the individual fading of each of the YSCs up to a common age 
of 12 Gyr.

\begin{figure} 
\includegraphics[width=\columnwidth]{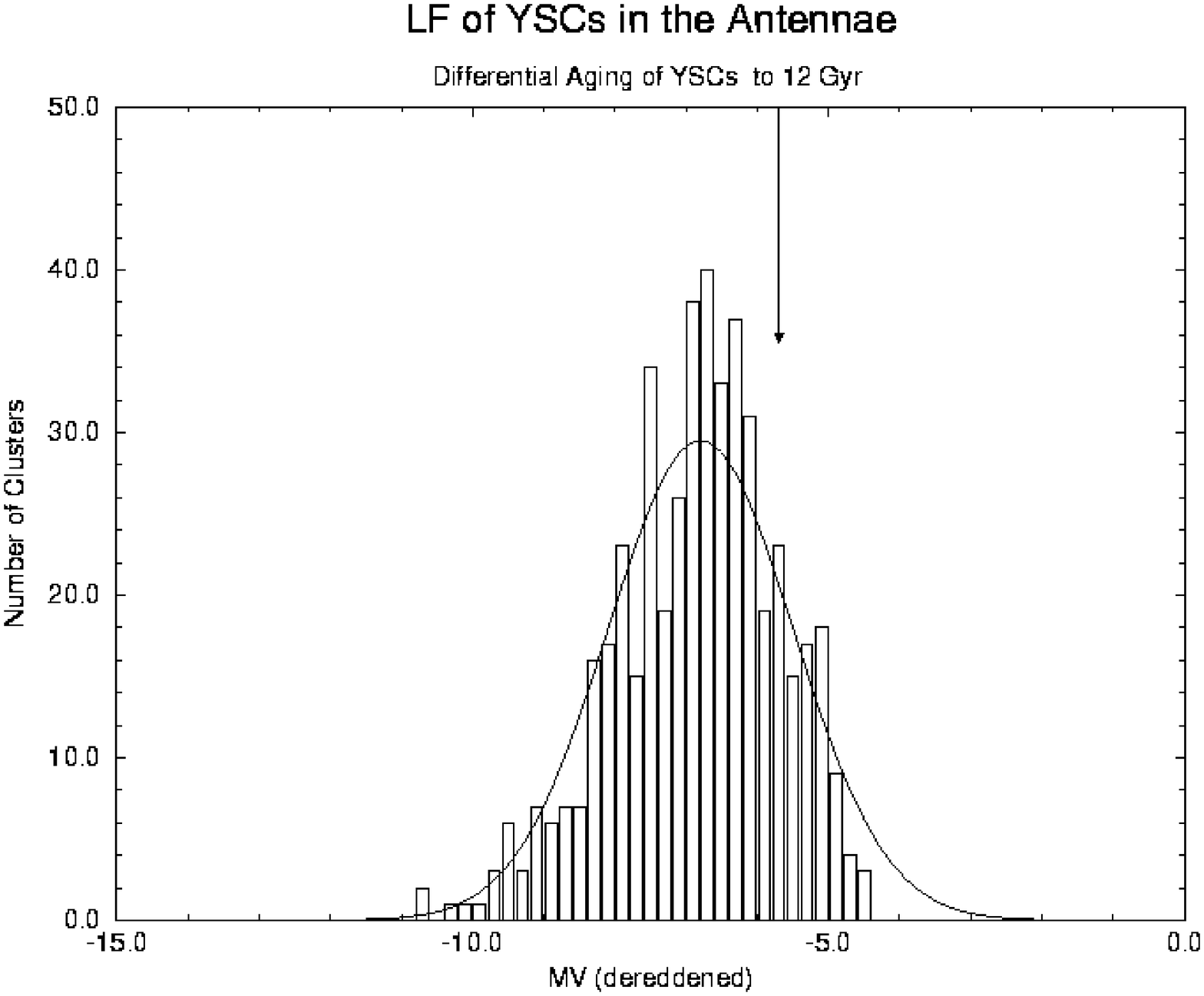}
\includegraphics[width=\columnwidth]{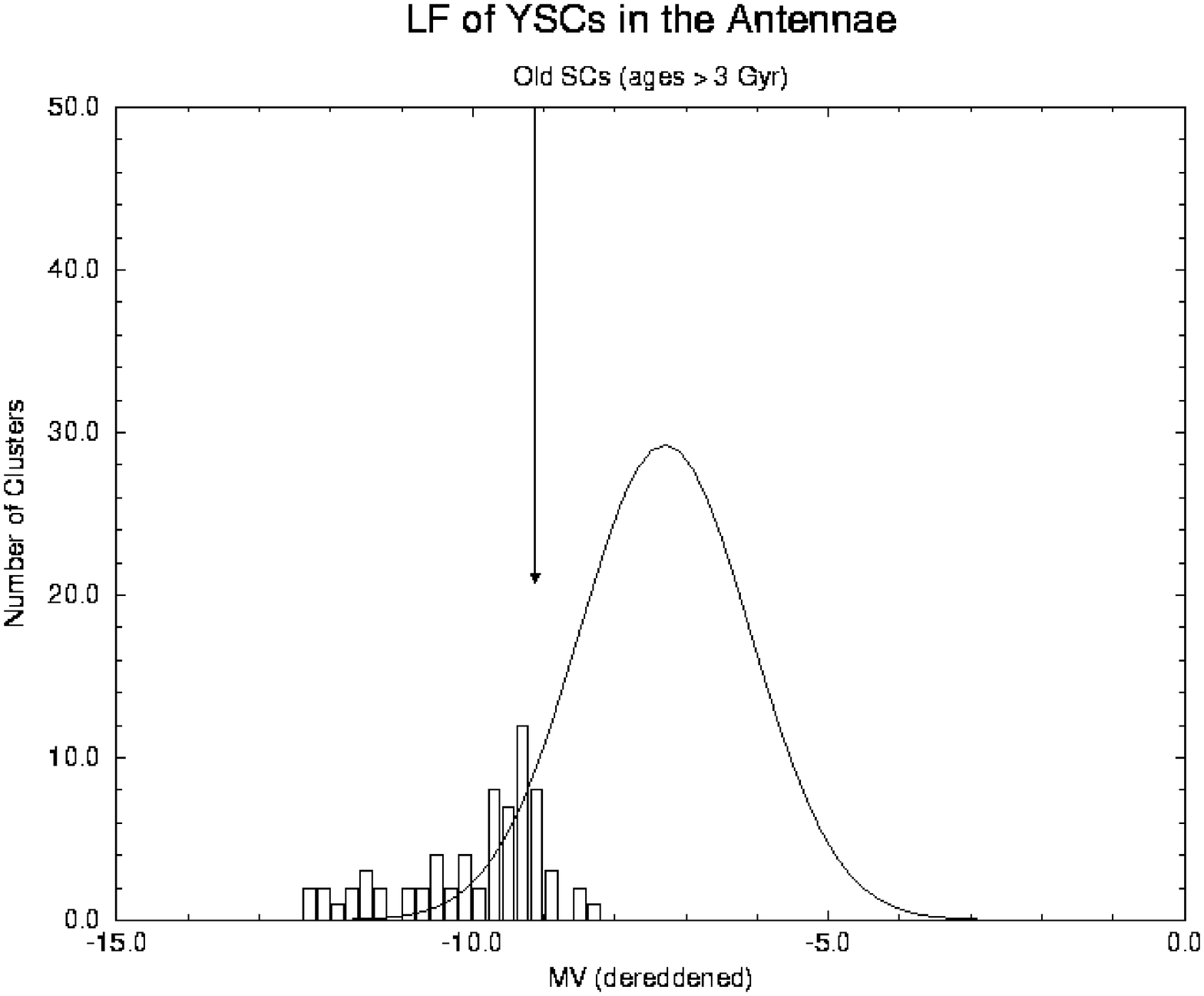}
 \caption{{\bf(6a)} LF of YSCs in the Antennae as calculated from individual 
 ages together with 
 the resulting individual fading until 12 Gyr for every cluster. A Gaussian 
 with $\langle {\rm M_{V_0}} \rangle =-6.9$ mag and $\sigma({\rm M_{V_0}})=1.3$ 
 mag is overplotted, 
 normalised to 
 the number of clusters in the histogram. {\bf(6b)} present dereddened LF 
 of the old GC population from the progenitor spirals together with a 
 Gaussian with $\langle {\rm M_{V_0}}\rangle =-7.3$ mag and 
 $\sigma({\rm M_{V_0}})=1.2$ mag 
 (Ashman \etal 1995), normalised to the total number of GCs in the Milky Way 
 and M31. Vertical arrows indicate the observational completeness limit.}
 \end{figure}

In Fig. 6a. we present the LF of the YSC population evolved until an age of 
12 Gyr, i.e. the aged LF of those 481 clusters with present ages $<3$ Gyr, 
for which we argued in Sect. 5. that they most probably 
formed in the ongoing galaxy merger. 
To estimate the position of the competeness limit in this plot, we take a YSC 
at the observational competeness limit given by WS95 which has the mean age of 
the YSCs and evolve it to 12 Gyr. It will thereby fade to ${\rm M_{V_0}} = -5.7$ 
mag which we call the evolved completeness limit. 

The LF of the YSCs aged to 12 Gyr 
clearly shows a turnover $\sim 1$ mag brighter than the evolved 
completeness limit. Overplotting a 
Gaussian with M$_{{\rm V}_0}=-6.9$ mag and $\sigma$(M$_{\rm V}) = 1.3$ mag, 
normalised to the number of clusters in our histogram, we find that the 
agreement is not too bad. The turnover occurs at an M$_{{\rm V}_0}$ fainter 
by $\sim 0.2$ mag than for typical GCSs. 

Ashman \etal 1995 show that the turn-over of the GCLF is metallicity dependent. 
While for spiral galaxies with their typical $\langle[{\rm Fe/H}]
\rangle_{\rm halo GCs} = -1.35$ they give a turnover around ${\rm M_{V_0}} = -7.3$ 
mag they 
find a turn-over fainter by $\sim 0.15$ mag for a typical elliptical with a 
characteristic metallicity of its GCS higher by 0.5 dex than in spirals, 
in agreement with FB95's models. We have argued that the metallicity of stars 
and clusters formed in the interaction-triggered starburst in the Antennae 
should be [Fe/H] $\geq -0.7$ and, by analogy to NGC 7252, probably around 
[Fe/H] $\sim -0.4$. This means that if some of the bright blue knots are 
young GCs a bimodal metallicity distribution as e.g. observed in the 
suspected merger remnants NGC 4472 and NGC 5128 (Harris \etal 1992, Zepf 
\& Ashman 1993, Ostrov \etal 1993) or in M87 (Elson \& Santiago 1996) 
should be expected to persist over a 
Hubble time  in the Antennae's GC system. The original, low-metallicity GC 
population of the progenitor spirals then should show a LF peaking at 
${\rm M_{V_0}} = -7.3$ while the higher metallicity secondary generation 
GCs will have a LF peaking at fainter magnitudes, e.g. for [Fe/H] $\gta -0.4$ at 
${\rm M_{V_0} \gta -7.0}$ (cf. Fig. 2 of Ashman \etal 1995). Thus, the higher 
metallicity of a secondary GC population might explain the turn-over around 
M$_{{\rm V}_0} \sim -6.9$ mag that is indicated in Fig. 6a. 

WS95 present in their Fig. 9 the present-day LF of the entire YSC sample, which 
clearly looks exponential up to the completeness limit. In Sect. 4 we showed 
that without taking age spread effects into account aging simply 
shifts the LF to fainter magnitudes without changing its shape. Comparison with 
our LF in Fig. 6a clearly shows the strong changes that the inclusion of the 
age spread among a YSC population induces on the shape of the LF during time 
evolution. In this respect we confirm and quantify Meurer's (1995) conjecture. 

A release of our simplistic assumption of a homogeneous metallicity for all $-$ 
older and younger $-$ YSCs in the sense that the youngest of them 
should be expected to have a higher metallicity than those which formed at the 
very beginning of the starburst (cf. Fig. 12 in Fritze $-$ v. Alvensleben \& 
Gerhard 1994) would 
bring along further repartition effects because the 
fading gets stronger as the metallicity of a star cluster increases: 
e.g. between ages of $10^7$ yr and 12 Gyr a YSC with 
${\rm Z = 1 \cdot 10^{-3}}$ would 
fade by 4.8 mag while with ${\rm Z = \half Z_{\odot}}$ it would fade by 
5.4 mag and with ${\rm Z = 2 \cdot Z_{\odot}}$ by as much as 5.6 mag (cf. Fig. 2 
in FB95). The most metal-poor YSCs are expected to be the oldest, i.e. already 
somewhat fainter that at birth. They will fade less during subsequent 
evolution while the most metal rich ones should be the youngest, i.e. 
among the brightest, and they will fade more than average. In this way, a 
metallicity spread would cause the LF to become narrower over a Hubble time. 
Our LF calculated with a single homogeneous metallicity is already well fit 
by a Gaussian with $\sigma$(M$_{\rm V}) = 1.3$ mag, the typical value for 
all known GCSs, so a strong metallicity 
spread among the Antennae's YSCs might lead to a very narrow final LF. 
Our metallicity prediction, however, needs 
confirmation by individual spectroscopy of some YSCs, which, at the same time, 
will give an impression of the possible importance of a scatter in metallicity 
or abundance ratios. 

The global detection limit given by WS95 does not exclude the possibility 
that in the most crowded and brightest regions un unknown number of YSCs brighter 
than this may be missed. Clusters in those regions  are expected to be perticularly 
young and will fade more than average, migrating to very faint bins in the aged 
LF of Fig. 6a. If their number were large enough, they might affect the shape 
of the LF. 

We caution that dynamical effects are still not included in our modelling. 
If they should be expected to further reshape the 
LF over a Hubble time $-$ beyond the photometric evolution effects discussed 
here $-$ will be investigated in Sect. 8. 

It came as a surprise to us that the evolved LF of {\bf all} the bright YSCs 
so closely resembles a typical GCLF. We started our analysis with 
the aim of finding out some selection criterion for a subsample of objects, 
as e.g. with small R$_{\rm eff}$ or high luminosity, that may evolve into an 
old GC population while we expected an unknown but $-$ in view of the relatively 
young age of the burst $-$ possibly important fraction of open clusters and 
associations to be present, too. Our results seem to indicate that either the 
bulk of the YSCs recently formed in NGC 4038/39 are indeed young GCs or else that 
from the range and distribution of integrated luminosities there are no strong 
intrinsic differences between young open and globular clusters {\bf and} that 
dynamical destruction does not significantly reshape the LF down to $\sim 1$ mag 
below the turnover. In this respect, it will be very interesting to extend the present analysis 
of the LF to fainter magnitudes using WFPC2 data. 

In Fig. 6b, we present the present-day, i.e. the observed and dereddened LF of 
the star cluster subsample with ages $\geq 3$ Gyr for which, in Sect. 5, we 
argued that they might well belong to the progenitor spirals' original GC 
population. It comprises 69 clusters, 55 of them brighter than the completeness 
limit of the observations corresponding to M$_{{\rm V}_0}=-9.1$ mag. We compare 
this bright end of the old star cluster LF to a Gaussian with 
M$_{{\rm V}_0}=-7.3$ mag and $\sigma = 1.2$ mag (Ashman \etal 1995), 
normalised to the total 
number of GCs in both the Milky Way and M31. While a different shape of their LF 
is, of course, not ruled out, the reasonable agreement at 
the high luminosity tail lends support to our conjecture that these red old 
clusters may belong to the original GC population. 
If the interacting galaxies NGC 4038 and 4039 together really had a number of 
GCs comparable to that of the Milky Way and M31, then the number of bright YSCs 
formed in the merger is of the same order as the number of GCs of the two 
spirals. This is the number ratio that is required if an elliptical galaxy 
with a typical GC frequency were to be formed from a merger of two spirals 
(Zepf \& Ashman 1993). Major 
uncertainties, however, come from the fact that the starburst and cluster 
formation may go on in the Antennae as well as from the unknown fraction of 
open clusters/associations among the YSC population. 

As seen in Fig. 2 of FB95, fading is strongest during the first Gyr 
($\Delta{\rm M_V} \sim 1$ mag from $2\cdot10^8$ to $1\cdot 10^9$ yr), weaker 
during intermediate stages ($\Delta{\rm M_V} \sim 0.42 
{{\rm mag}\over{\rm Gyr}}$ for ages 1 $-$ 6 Gyr), 
and very weak at old ages ($\Delta{\rm M_V} \sim 0.075 
{{\rm mag}\over{\rm Gyr}}$ for 8 $-$ 16 Gyr). Thus, if we evolve the 
LF until 16 Gyr instead of 12 as in Fig. 6a., its shape does not 
change any more, it will only be shifted by $\Delta{\rm M_V} \sim 0.3$ to 
slightly fainter magnitudes. 

Age spread effects will, of course, reshape not only the LF, but also the 
colour distribution of a YSC system. Over a Hubble time, the  internal dust 
extinction in the Antennae may also be expected to change a lot, as the 
starbursts consumes and/or blows out the gas and dust now observed. 
The bimodal 
metallicity distribution we predict for the Antennae's GC system will result in 
a bimodal colour distribution similar to the one found by Whitmore \etal (1995) 
and Elson \& Santiago (1996) for the M87 GC system.

\section{Discussion}
\subsection{Comparison of young and old star cluster properties}

Here we compare average properties of the red (${\rm (V-I)}_0 \geq 0.95$) and blue 
(${\rm (V-I)}_0 < 0.95$) star cluster subsamples. We argued that the red subsample 
may mainly consist of the brightest of the original spirals' GCS 
while we expect the blue subsample to contain some mixture of young open clusters, 
associations and globular clusters.

\begin{table}
\caption{Comparison of old and young star cluster subsamples.}
\halign{\hfil#\hfil&\hfil#\hfil&\hfil#\hfil \cr
\noalign{\medskip\hrule\medskip} \cr
 & ${\rm (V-I)}_0 \geq 0.95$ &  ${\rm (V-I)}_0 < 0.95$ \cr
\noalign{\medskip\hrule\medskip} \cr
 ${\rm N_{obj}}$  & 69  & 481  \cr
 $\langle {\rm R_{gc}} \rangle$  & $3.92 \pm 2.01$ kpc & $3.42 \pm 1.57$ kpc \cr
 $\langle {\rm V} \rangle$ & $~22.05 \pm 0.99$ mag ~ & $~21.36 \pm 1.05$ mag ~ \cr
 $~\langle {\rm V-I}  \rangle ~$ & $~1.61 \pm 0.32$  & $~0.61 \pm 0.34$ \cr
 $\langle {\rm U-V}  \rangle$  & $~~~ - ~~~$  & $-0.70 \pm 0.25$ \cr
 $\langle {\rm R_{eff}} \rangle$ & $12.04 \pm 6.95$ pc & $12.41 \pm 6.22$ pc \cr
\noalign{\medskip\hrule\medskip}}
\end{table}

Table 2 summarises average quantities of the two subsamples with 
${\rm (V-I)_0} < 0.95$ and ${\rm (V-I)_0} \geq 0.95$. 
For 164 clusters, no ${\rm (V-I)}$ 
observations are available. It is seen that the 69 red 
clusters populate an area within NGC 4038/39 somewhat more extended than the 
481 blue ones. If dust were somehow concentrated to the center, 
the redder subsample should be less affected by dust than the bluer one, 
thus increasing the colour difference. 

Plotted separately as projected on the sky, the blue YSCs and the red GCs show 
very different spatial distributions. While the blue clusters tightly trace the 
tidal structure as seen on WS95's HST image of the Antennae, the red clusters 
show a more spherically symmetric distribution as expected for the original 
GC population.

The redder clusters are fainter by $\sim 0.7$ mag with a scatter slightly 
smaller that the blue ones. If the difference in $\langle {\rm V-I} \rangle$ of 
1 mag were due to stronger than average internal dust reddening the red clusters 
should be fainter than the blue ones by as much as 1.7 mag, on average. 

Their $\langle {\rm V-I} \rangle_0 =1.31 \pm 0.32$ 
corresponds to ages from 4 $-$ 15 Gyr for Z=0.01 and from 1 $-$ 10 Gyr for 
Z=0.04, respectively, clusters with ${\rm Z < 0.01}$ in our models do not reach 
colours as red as ${\rm (V-I)}_0 = 1.3$ until 15 Gyr, values ${\rm (V-I)}_0 > 1.6$ is not even 
reached by our model clusters with ${\rm Z = 0.04}$. So we suspect, that these very 
red clusters are affected by stronger than average dust reddening. 
The average $\langle {\rm V-I} \rangle_0$ of the 
bluer clusters is bluer by 1 mag than that of the redder ones 
with a larger scatter and indicates a mean age 
of $1 \cdot 10^8$ yr with a range from $5 \cdot 10^6$ to $5 \cdot 10^8$ yr for 
Z = 0.01.  
The mean dereddened luminosity of the red subsample is $\rm \langle {\rm M_{V_0}} 
\rangle = -9.87 \pm 0.99$, so it is only the brightest of the 
really old GCs from the 
parent galaxies of the merging system, which are seen in the red subsample.

From our models, the age difference between the subsamples corresponds 
to a fading by $\sim 4.2$ mag if they have the same metallicity and of 4.6 mag 
if the redder ones have ${\rm Z = 1 \cdot 10^{-3}}$. If the redder ones are older and 
fainter, however, only the very brightest of them can be observed, so that their 
mean brightness is overestimated. On the other hand, this 
difference may aresult if chances to survive a Hubble 
time were larger for brighter and more massive clusters (cf. Sect.8).

\subsection{Effective radii of GCs and YSCs}

The mean effective radius of the red subsample is slightly smaller than that of 
the bluer clusters. If 
they are really old this can be understood in terms of 
a preferential destruction of large R$_{\rm eff}$ clusters strong enough to 
overcome the increase in R$_{\rm eff}$ caused by internal stellar mass 
loss ($\sim 20 \%$ in R$_{\rm eff}$) and mass loss through tidal stripping (cf. 
Sect. 8). 
The difference in  R$_{\rm eff}$ is surprisingly small, however, in view of our 
expectation that the old star cluster population should consist of GCs while the 
blue and young star population might comprise open clusters and 
OB associations, as well. 
The distribution of effective radii of the YSC subsample has a clear maximum 
around 8 pc with a tail extending to $\sim 32$ pc, that of the old GC sample is 
somewhat flatter but otherwise very similar.

The breakdown of the age distribution into clusters with small and large 
${\rm R_{eff}}$ in Figs 5b and 5c shows that while of the 201 clusters with 
${\rm R_{eff}} \leq 10$ pc 11\% belong to the old and 89\% to the young 
population, the respective fractions are 13\% and 87\% for the 349 clusters with 
${\rm R_{eff}} > 10$ pc. It is clear that the old objects must almost all be 
globulars, even those with ${\rm R_{eff}} > 10$ pc, 
except for a small number of very old open clusters as reported for the 
Milky Way by Friel (1995). 
Among the old GC population the 
number ratio of objects with ${\rm R_{eff}} \leq 10$ pc to those with 
${\rm R_{eff}} > 10$ pc is 23/46 while among the YSC population it is 178/303. 
This corresponds to a fraction of objects with ${\rm R_{eff}} \leq 10$ pc of 
33\% among the old GC population, while this fraction  amounts to 37\% in the 
YSC population. 

{\bf To summarize}, the red old GC population is slightly more extended than 
the population of young blue star cluster forming in the merger-induced 
starburst in the Antennae. While the blue YSCs trace the tidal structure, the 
red clusters rather show a spherically symmetric distribution. 
Interestingly, the mean as well as the distribution of {\bf effective radii are 
not significantly different for the bright end of the old GC population and 
for the less than 1 Gyr old young star cluster 
population} which might be expected to also comprise an unknown fraction of open 
clusters and OB associations together with young GCs. 

\subsection{Luminosity evolution of compact and extended YSCs}

In Figs. 7a and b, we present the LFs of the YSCs (age $< 3$ Gyr) with 
${\rm R_{eff}} \leq 10$ pc and ${\rm R_{eff}} > 10$ pc, respectively, both 
differentially aged to 12 Gyr. No significant difference comparable to the one 
discussed in Sect. 4 is visible any more. 
Strikingly and at variance with our simplified analysis in Sect. 4., {\bf both 
the differentially aged LFs of subsamples with R$_{\rm eff} \leq 10$ pc and 
R$_{\rm eff} > 10$ pc 
now do show a turnover at M$_{{\rm V}_0} \sim -6.9$ mag}, as seen in 
Figs. 7a, b.

\begin{figure} 
\includegraphics[width=\columnwidth]{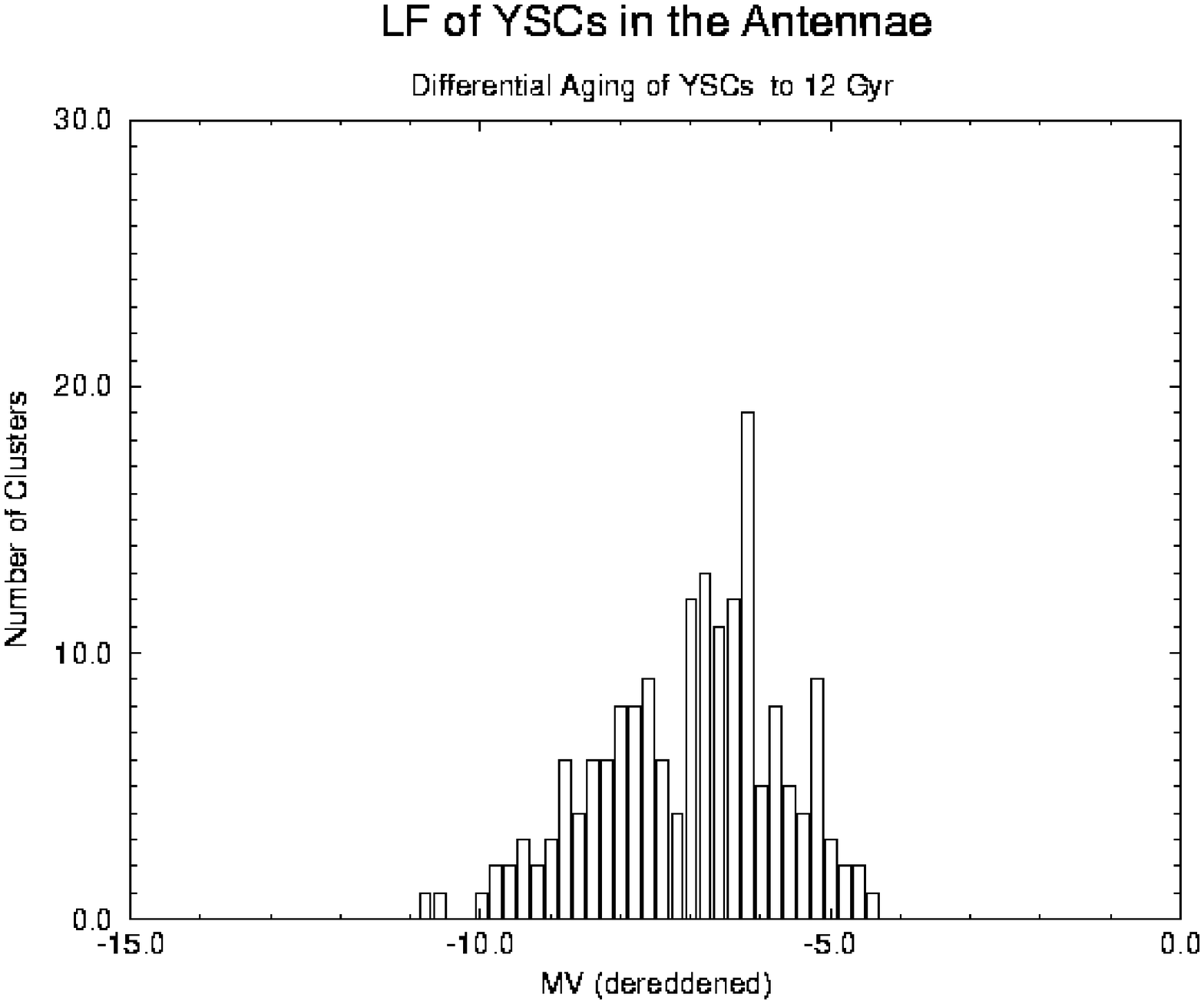}
\includegraphics[width=\columnwidth]{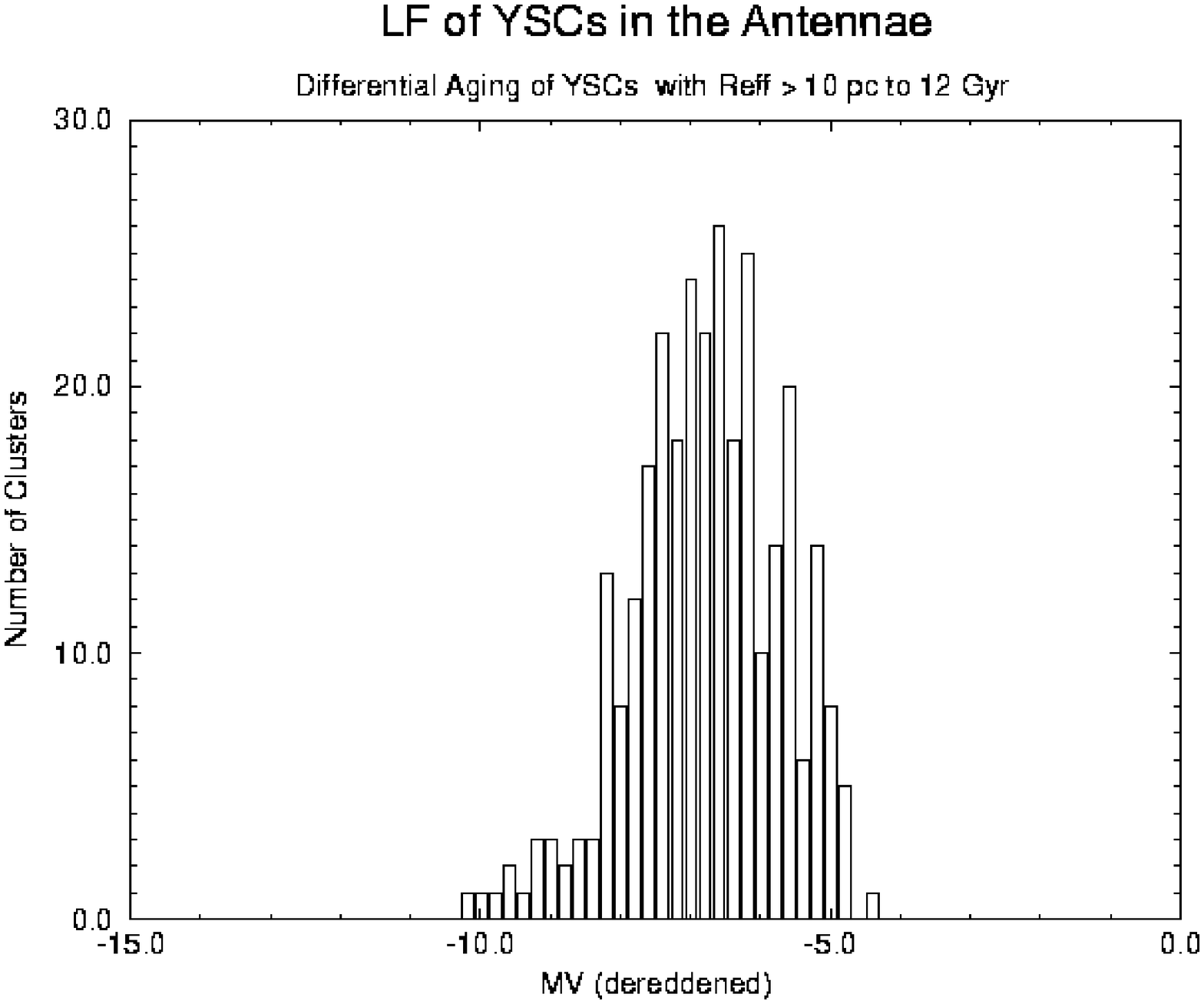}
 \caption{Differentially aged LFs of YSCs in the Antennae at 12 Gyr.  
 {\bf(7a)} clusters with
 R$_{\rm eff} \leq 10$ pc and {\bf(7b)} clusters with R$_{\rm eff} > 10$ pc.}
 \end{figure}

So, while no doubt some open clusters may be among the YSC population, preferentially 
within the large ${\rm R_{eff}}$ subsample, it might well be that the bulk 
of the YSC population are young GCs.

\subsection{Subdivision with respect to luminosity}
There is another plausible criterion to subdivide the YSC sample of WS95 into a 
subclass that might be expected to contain a significant number of young GCs as 
opposed to the rest of the sample that could possibly be dominated by open 
clusters and associations, and this is luminosity. 
One might conjecture that the most luminous objects might be young globulars 
while the fainter ones could also be 
open clusters/associations. To explore this hypothesis we subdivide the YSC 
sample of WS95 into subsamples of bright and faint clusters with a limiting 
${\rm M_V} = -9.5$, corresponding to a dereddened ${\rm M_{V_0}} = -10.0$. 
Table 3 presents the average properties of the faint and bright subsamples.
Our choice of the limiting magnitude makes both subsamples contain comparable 
numbers of objects.

\begin{table}
\caption{Comparison of young star cluster subsamples.}
\halign{\hfil#\hfil&\hfil#\hfil&\hfil#\hfil \cr
\noalign{\medskip\hrule\medskip} \cr
 & ${\rm M_V} > -9.5$ mag &  ${\rm M_V} \leq -9.5$ mag \cr
\noalign{\medskip\hrule\medskip} \cr
 ${\rm N_{obj}}$  & 358  & 356  \cr
 $\langle {\rm R_{gc}} \rangle$  & $3.79 \pm 1.73$ kpc & $3.29 \pm 1.51$ kpc \cr
 $\langle {\rm V} \rangle$ & $~22.50 \pm 0.36$ mag ~ & $~20.85 \pm 0.88$ mag ~ \cr
 $~\langle {\rm V-I}  \rangle ~$ & $~0.93 \pm 0.48$  & $~0.61 \pm 0.43$ \cr
 $\langle {\rm U-V}  \rangle$  & $-$-  & $-0.70 \pm 0.25$ \cr
 $\langle {\rm R_{eff}} \rangle$ & $7.88 \pm 3.98$ pc & $8.94 \pm 4.57$ pc \cr
\noalign{\medskip\hrule\medskip}}
\end{table}

The fainter clusters, on average, seem to have larger galactocentric distances. 
It is not clear, however, if this is a real effect or due to the fact that low luminosity 
clusters are harder to detect near the center. 
The mean V $-$ magnitude of the faint subsample is $\sim 1.7$ mag lower than the 
$\langle{\rm V}\rangle$ of the 
brighter ones, the rms scatter is less than half that of the brighter ones. 
Faint clusters are redder by 0.3 mag in ${\rm (V-I)}$. Unfortunately, no 
${\rm (U-V)}$  colours are 
available for the faint subpopulation. 
It should be noted that the differences in V and ${\rm (V-I)}$ cannot be explained by 
reddening differences but may be understood in terms of age differences (see below). 
Finally, the mean effective radius of 
the fainter 
subpopulation is smaller by $\sim 1$ pc and the rms scatter is smaller, too, 
than the respective values of the 
bright subsample.  All differences between the bright and faint subsamples 
increase if we chose a brighter limiting magnitude for our subdivision. 

The bright clusters have a mean age of their young population of 
$1.5 \cdot 10^8$ yr, younger by almost a factor of 2 than that of
$2.7 \cdot 10^8$ yr for the fainter ones. 
In the age distribution of the bright subsample the vast 
majority of clusters is young, only 24/331 are 3 $-$ 12 Gyr old. 
The age distribution of the faint clusters shows 45/219 clusters to be 
old (3 $-$ 12 Gyr) and 175/219 to be young (0 $-$ 3 Gyr). In Sect. 5, we argued 
that because of inhomogeneous metal and dust distributions probably 
most clusters with ages $< 3$ Gyr are as young as $\lta 4 \cdot 10^8$ yr and 
most clusters with ages $> 3$ Gyr are as old as 12 Gyr. 
After differentially aging all clusters from their present individual ages to 
an assumed final age of 12 Gyr, however, the brighter subsample will keep a 
brighter final luminosity $\langle{\rm M_{V_0}}\rangle = -7.4$ mag than the 
fainter subsample which will end up with $\langle{\rm M_{V_0}}\rangle = -6.7$ 
mag. As shown in Fig. 8, 
after 12 Gyr, the LF for the bright subsample clearly shows a turn-over 
definitely brighter than the completeness limit and does closely resemble 
the Gaussian LF of GCs with typical parameters $\langle {\rm M_{V_0}} 
\rangle = -7.1$ mag, $\sigma = 1.3$ mag rather than the exponentially 
increasing LF of open clusters. The fact that this turn-over is slightly 
brighter than the one we expect on the basis of our metal prediction $-$ 
if it is true $-$ might indicate that a limiting magnitude ${\rm M_{V_0}} = -10$ 
is still so bright as to exclude part of the young GC population. 
The LF of the faint clusters contains an important fraction of 
$\sim 12$ Gyr old clusters. If these are left aside when the LF is evolved 
over a Hubble time, we again observe a turn-over, however only slightly before 
the completeness limit is reached, i.e. at $\langle {\rm M_{V_0}} \rangle = 
-6.2$ mag.

\begin{figure} 
\includegraphics[width=\columnwidth]{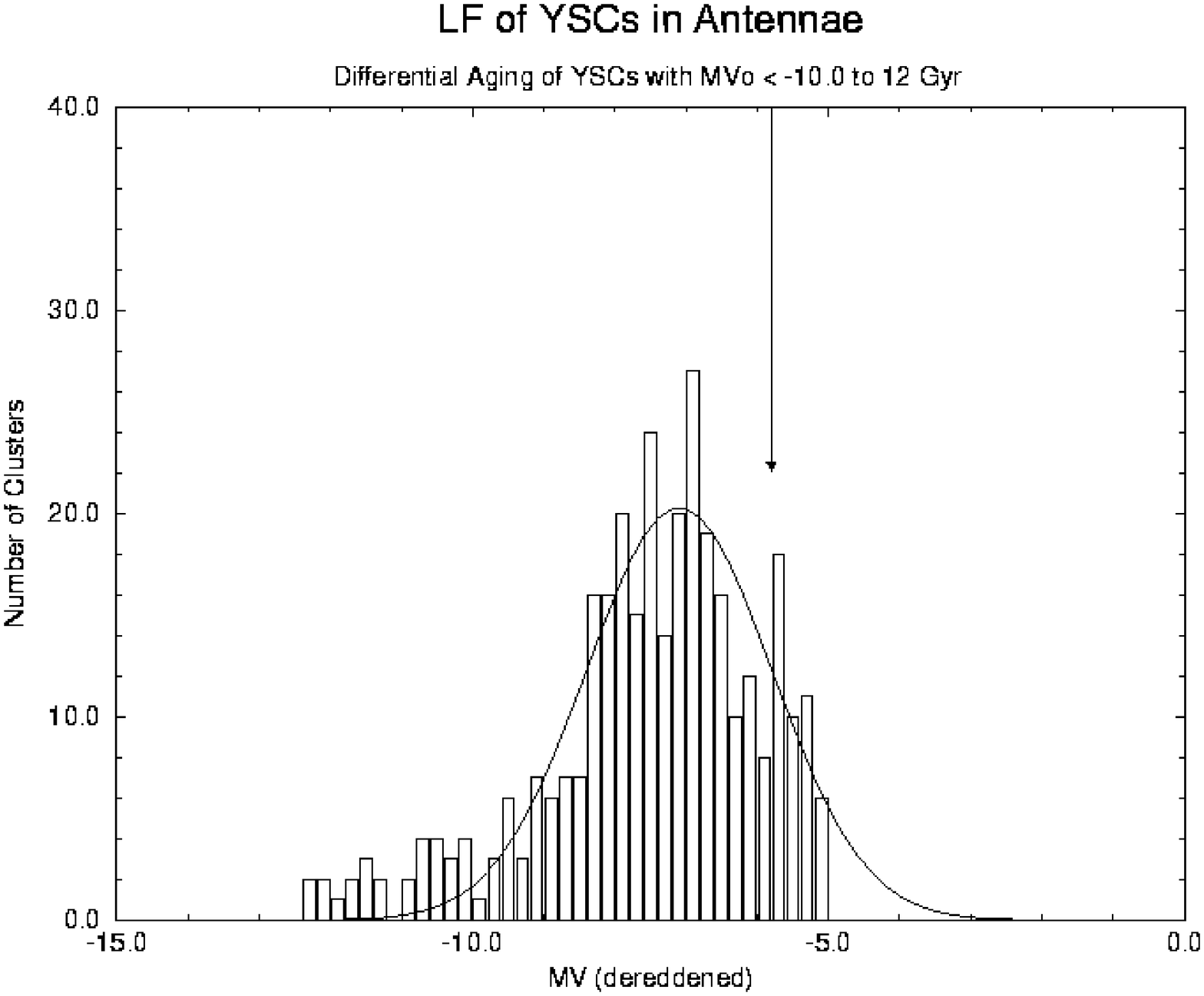}
 \caption{LF of bright YSCs with ${\rm M_{V_0}} \leq 
 -10.0$ mag differentially aged to 12 Gyr. Overplotted is a Gaussian with 
 $\langle{\rm M_{V_0}} \rangle = -7.1$ mag, $\sigma = 1.3$ mag, normalised to 
 the number of YSCs in the histogram.} 
 \end{figure}

{\bf To summarise:} the faint cluster subsample contains a $\sim 20$\% 
proportion of $\sim 12$ Gyr old GCs, it is a little less concentrated to the 
center on average than the bright subsample. It has marginally smaller 
$\langle{\rm R_{eff}}\rangle$ than the bright one. Unless errors of the 
effective radii are systematically 
different for the bright and faint subsamples, the larger $\langle
{\rm R_{eff}}\rangle$ 
of the bright sample seem to reflect an initial correlation of larger 
radii for higher luminosity clusters that overcomes the age effect of 
increasing the radius due to mass loss. The LF of the subsample of clusters 
brighter than ${\rm M_{V_0}} = -10.0$ mag evolved over 12 Gyr closely 
agrees with a typical GC system's LF ($\langle {\rm M_{V_0}} \rangle = -7.1$ 
mag, $\sigma = 1.3$ mag), provided age spread effects are properly accounted for.

\subsection{Need for Future Observations}

All the data discussed here were taken by WS95 with 
HST WFPC1, i.e. before the refurbishment mission. 
The Antennae are close enough to even identify the brightest members 
of the original GC population on these aberrated images. 
WS95 expect reobservations with WFPC2 to go deeper by 2$-$3 mag. If this 
is true and the completeness limit can really be pushed down that far we might 
be able to observe beyond the turn-over of the original GC population. Then 
both the 
old GC population and the YSC population can directly be compared. It would 
be very 
interesting to have spectroscopy for some of the clusters 
to settle their metallicity range 
and to allow for better age-dating. At the same time, YSC spectra will provide 
information about the local dust distribution and, if spectral resolution 
allows, also about the kinematics of the YSC system. 

The very youngest star clusters will give us the chance to directly study 
the upper IMF.

\section{Dynamical effects on the GCLF}

It is generally accepted that an old GCS, as e.g. observed in the Milky Way, 
only encompasses ``the hardiest survivors of a larger original population'' 
(Harris 1991). 

Fall \& Rees (1977) were the first to discuss the erodive effect on a GCS 
through its environment. Ostriker (1988) lists a number of $-$ partly 
interrelated $-$ processes, that modify star cluster properties and decimate 
the original cluster population.  
Dynamical effects are conveniently distinguished into internal dynamical effects 
due to stellar mass loss and external dynamical effects from the parent galaxy 
potential.

\subsection{External dynamical effects}

The most important external dynamical effect for clusters in the Antennae is 
probably tidal shocking of 
clusters near the center and of clusters on highly eccentric orbits that 
pass close to the high density central parts of the merging system, in 
particular, if the Antennae is to evolve into a high central density elliptical 
or S0 - like merger remnant. Tidal shocking leads to evaporation and is 
strongest for low concentration clusters, which, in turn, tend to be those with 
lower luminosity (Djorgovski 1991). Dynamical friction,as well, may remove 
clusters near the center, and this process preferentially acts on the massive 
ones. 

For GC populations in non-interacting 
galaxies, these effects were studied by Chernoff \& Weinberg (1990), their 
results are largely confirmed by the independent and more realistic approach 
of Fukushige \& Heggie (1995). 

Unfortunately, it is the concentration parameter 

\centerline{c := Log R$_{\rm T}$/R$_{\rm C}$ (= Log R$_{\rm T}$/R$_{\rm eff}$ before 
core collapse)} 
\noindent
that $-$ besides the orbital parameters $-$ plays the key role 
for the question of survival or destruction of a star cluster. Due to the fact 
that tidal radii are not accessible observationally the concentration parameters 
of YSCs as well as those of the old GCs in merger remnants like the Antennae remain 
unknown. An estimate of c from correlations to other quantities (eg. luminosity) 
as found in observations on {\bf old} GCSs does not seem reasonable since it is 
not clear to what extent these 
correlations reflect intrinsic trends already present when these GCSs were young 
or else are, themselves, the result of dynamical effects. 

Recent semianalytic modelling of the Galactic GCS by Vesperini (1997) yields 
the interesting result that $-$ provided ``the mass function is initially taken to 
be a log-normal distribution similar to the one currently observed in our Galaxy, 
its shape is not significantly altered during the entire evolution even though 
a significant number of clusters are disrupted before one Hubble time''. 
If the same were true in the case of the merging pair of the Antennae galaxies, it 
would mean that the LF should not change its shape $-$ 
beyond what we found from the 
proper consideration of age spread effects $-$ during 12 Gyr of dynamical 
evolution allthough a significant number of clusters will be destroyed. 

The process of referring the LF to a uniform age of the YSC system, as we have 
done in Fig. 6a properly accounting for age differences in the presently observed 
YSC population mimics a transition from a LF to a cluster mass function since at 
fixed age models indicate a fixed M/L for all YSCs. Fig. 6a thus shows that the 
mass function of the YSC system currently forming in NGC 4038/39 is similar to a 
Gaussian, which, according to Vesperini's results seems to represent a sort of 
quasi-equilibrium distribution capable of surviving a Hubble time with its shape 
and parameters conserved.  

All these external dynamical effects are already difficult to model in a 
galaxy for which the potential is comparatively easy to describe and not 
variable in time. It seems extremely difficult, however, to model them in a 
quantitative way in an ongoing merger like the Antennae. Kinematic information 
from YSC spectroscopy with 10 m class telescopes together with 
a detailed dynamical modelling of the interaction process including gas dynamics 
and a SF criterion may bring further insight. 

\subsection{Internal dynamical effects}

Internal dynamical effects are easier to estimate since they only weakly 
depend on the external tidal field and they are important since the dynamical 
evolution of young GCs is dominated by adiabatic mass loss due to stellar 
evolution (Chernoff \& Weinberg 1990)

In addition to the photometric evolution, our models also give the mass ejection 
rates from stars as a 
function of time including stellar winds, SNe, and PNe. Thus, they allow to 
directly follow the time 
evolution of the total stellar mass of a star cluster. In Fig. 9, we present 
the time evolution of 
the relative fraction of stellar mass lost from a star cluster with a Scalo 
(1986) or Salpeter (1955) IMF, a lower mass limit of 0.1 M$_{\odot}$ and various 
upper mass limits and initial metallicities.

\begin{figure} 
\includegraphics[width=\columnwidth]{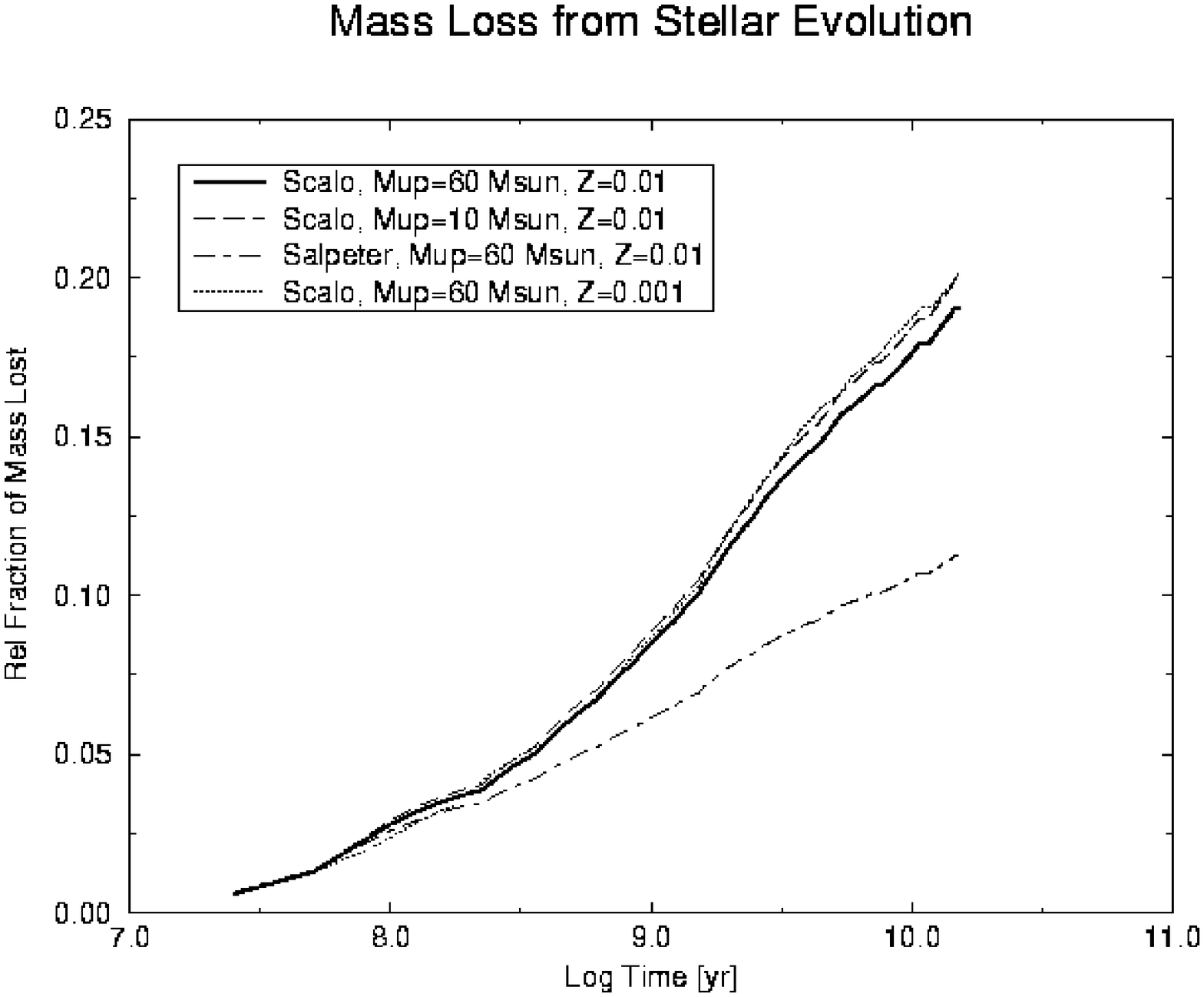}
 \caption{Time evolution of the relative fraction of mass lost from a star 
 cluster as calculated from our models for various IMFs, upper mass limits, and 
 metallicities.}
 \end{figure}

For the case of a Scalo IMF, the time evolution of the mass 
loss rate can be well approximated by two linear regimes: a relative mass loss 
of about 2\% per $10^8$ yr 
during the first $3 \cdot 10^8$ yr, and of 1\% of the total mass per Gyr over 
the rest of the Hubble time.
Within the first $3 \cdot 10^8$ yr, $\sim 28 \%$ of the total mass loss occurs, 
during the first Gyr $\sim 50 \%$ 
of the final mass loss is accomplished. These results confirm the idea that most 
of the young GCs that are 
going to be destroyed over a Hubble time will not even survive their first Gyr 
(Freeman 1995, priv. comm.).

For comparison, we have also calculated models with upper mass limits of 10 and 
60 $M_{\odot}$ and we 
do not find any significant differences in the mass loss rates. Also, differences 
for clusters of various initial 
metallicities are small (cf. Fig. 9).

As compared to a Scalo IMF, a Salpeter IMF with the same normalisation to total 
mass contains more stars above 6.5 ${\rm M_{\odot}}$ and below 0.5 
${\rm M_{\odot}}$. The latter ones, however, do not contribute to mass loss as 
their lifetimes are longer than a Hubble time. At the same time, a Salpeter IMF 
contains less stars in the range 0.5 $-$ 6.5 ${\rm M_{\odot}}$. For both types 
of IMF the total mass loss over a Hubble time is dominated by stars 
$< 6.5 {\rm M_{\odot}}$, more massive stars only account for $\sim 1.5$ \% 
of the total mass loss. This explains why for a Salpeter IMF with same lower 
and upper mass limits, the total mass loss is less than for a Scalo IMF. 
By the end of the first Gyr, a cluster with Scalo IMF has lost $\sim 8$ \% of 
its mass, with Salpeter IMF only $\sim 6$ \%. After 3 Gyr a Scalo cluster has 
lost $\sim 13$ \% and a Salpeter cluster $\lta 9$ \% of its mass through 
internal stellar evolutionary processes. 

To maintain virial equilibrium, a star cluster's effective radius increases 
by the same percentage by which its mass decreases. The core collapse and 
subsequent reexpansion processes (cf. eg. Bettwieser \& Fritze 1984) do not 
significantly affect the effective radius of a star cluster, they merely 
decouple the core radius from the effective radius. 

\subsection{Observational prospects}
If, as suggested by the similarity of GCSs in galaxies of very different 
Hubble types, luminosities and metallicities, the formation process of GCs 
does not strongly depend on details of environmental conditions, comparison 
of YSC systems in starbursts of different ages may offer the possibility to 
follow the time evolution 
of YSC properties and to study from an 
observational point of view the influence of dynamical effects on a YSC 
population which otherwise are difficult to model in the case of ongoing or 
recent galaxy mergers.

\section{Conclusions}
Using our method of evolutionary synthesis for various metallicities we present 
a first analysis of WS95's WFPC1 data on bright star clusters in the ongoing 
merger-induced starburst in NGC 4038/39. Assuming a metallicity Z $\sim 0.01$ 
on the basis of the progenitor spirals' ISM properties and applying a uniform 
reddening as given by WS95 we age-date the bright cluster population from their 
(V$-$I) colors and, as far as available, also from their (U$-$V). 
It turns out that in addition to a large population of young clusters with a 
mean age of $2 \cdot 10^8$ yr (consistent with  the dynamical time since 
pericenter) part of the original spirals' old GC population is also observed. 
A key question with far-reaching consequences as to the origin of elliptical 
galaxies is whether there are a significant fraction of young GCs among the 
YSC population. Two basic 
properties discriminate open clusters/OB associations from GCs in our 
Galaxy and others: the concentration parameter c = log (${\rm R_T/R_{eff}}$) and 
the LF which, in contrast to that for an open cluster system, is Gaussian 
for {\bf old} GCSs. Tidal radii and, consequently, concentration parameters not 
being accessible to observations in distant galaxies we examine the LFs 
of cluster subsamples with large and small effective radii.

In a first step, using a common mean age for all young clusters and a 
corresponding uniform fading to an age of $\sim 12$ Gyr we find that while 
the LF for extended clusters at 12 Gyr is definitely not Gaussian, that for the 
low R$_{{\rm eff}}$ clusters may well contain a Gaussian  (= GC) 
subcomponent together with a strong overpopulation of the faint bins, which 
themselves, however, might be expected to be severely depopulated over a Hubble 
time by dynamical effects not included in our models. 

Since for an ongoing starburst the age spread among YSCs may be of the same 
order as their ages, age spread effects are expected to reshape the LF. Clusters 
from the bright end tend to be younger on average and fade more than clusters 
from the faint end. We therefore, in a second step, model the individual fading 
consistent with individual ages of the YSCs as derived from their ${\rm (V-I)}$ 
and ${\rm (U-V)}$ colours, and we follow the LF changing its 
shape over a Hubble time. 

Surprisingly, accounting for these age spread effcets, we find the final LFs of 
large {\bf and} small R$_{\rm{eff}}$ cluster subsamples not to be significantly 
different any more. Instead, the LF of {\bf all} YSCs evolved to a common age 
of 12 Gyr is well compatible with a ``normal'' GCLF. Its turn-over occurs at 
$\langle {\rm M_{V_0}} 
\rangle \sim -6.9$ mag, i.e. slightly fainter than the average value $\langle 
{\rm M_{V_0}} \rangle \sim -7.1$ mag for 16 galaxies. This difference is readily 
explained in terms of a higher metallicity of the secondary cluster population. 

The number of old GCs from the spiral progenitors is consistent with the number 
of bright GCs expected if the progenitors had GCSs similar to the ones in 
the Milky Way and M31. 

Strikingly, neither the mean nor the distribution of effective radii is 
significantly different for the old GC sample and for the YSC sample. 
On the basis of these WFPC1 data we tentatively conclude that the bulk of the 
YSC population detected in the Antennae might well be young GCs and that the 
open clusters/associations probably also present among the YSCs do not seem to 
systematically differ from young GCs in terms of R$_{\rm{eff}}$. 
We are looking foreward to repeat this kind of analysis on WFPC2 data which 
may reach close to the old GCS's turn-over, reveal a number of fainter young 
objects, and will allow for more precise and definite conclusions. 

Dynamical effects that eventually might further reshape the LF over a Hubble 
time are discussed. 
Referring the YSCs' luminosities to a uniform age allows to recover the 
intrinsic mass function of the YSC system. This mass function seems to be 
log-normal which, according to Vesperini (1997), represents a 
quasi-equilibrium distribution that is going to be preserved in shape though 
not in number of clusters over a Hubble time of dynamical evolution. 

Dynamical effects, however, are extremely difficult to model in detail in 
an ongoing merger. Comparison of YSC populations in mergers/starbursts of 
various ages seems a promising tool in an attempt to understand these effects 
from an observational side.

\vskip 1 cm 

{\sl Acknowledgements.} 
I am deeply indebted to B. Whitmore \& F. Schweizer for valuable discussions, 
encouragement and 
for sending us their star cluster data in machine readable form. 
I am grateful to Ken Freeman, Tom Richtler, and Andreas Burkert for interesting 
discussions on dynamical aspects. 
I wish to thank Prof. Appenzeller and all the collegues from the 
Landessternwarte Heidelberg for their warm hospitality during a 3 months stay, 
when this projected was begun. 
My deep thanks go to the referee, G. Meurer, for his very detailed and constructive 
suggestions that greatly improved the paper. 
I gratefully acknowledge financial support from the SFB Galaxienentwicklung 
in Heidelberg and 
through a Habilitationsstipendium from the Deutsche Forschungsgemeinschaft 
under grant Fr 916/2-1 in G\"ottingen.


%

\end{document}